\documentclass[12pt]{article}

\usepackage[margin=1in]{geometry}
\usepackage{setspace}
\usepackage{microtype}
\usepackage[T1]{fontenc}
\usepackage{lmodern}
\usepackage{amsmath,amssymb}
\usepackage{booktabs,array}
\usepackage{graphicx}
\usepackage{pdflscape}
\graphicspath{{./}}
\usepackage{tikz}
\usetikzlibrary{arrows.meta,positioning,calc}
\usepackage{float}
\usepackage{caption}
\usepackage[hidelinks,colorlinks=true,citecolor=blue,linkcolor=blue,urlcolor=blue]{hyperref}
\usepackage[round,authoryear]{natbib}
\bibpunct[, ]{(}{)}{,}{a}{}{,}
\usepackage{url}

\renewcommand{\topfraction}{0.82}
\renewcommand{\bottomfraction}{0.62}
\renewcommand{\textfraction}{0.18}
\renewcommand{\floatpagefraction}{0.72}
\newcommand{\appref}[1]{\autoref{#1}}
\newcommand{\apprefrange}[2]{%
  \hyperref[#1]{Sections~\ref*{#1}}--\hyperref[#2]{\ref*{#2}}}
\newcommand{\exhibitnotes}[1]{%
  \par\vspace{6pt}
  \begin{minipage}{0.96\linewidth}
  \footnotesize\setlength{\parindent}{0pt}%
  \textit{Notes:} #1
  \end{minipage}%
}
\newcommand{\panelN}{45{,}386}
\usepackage{mathpazo}
\usepackage{microtype}
\usepackage{sectsty}
\usepackage{setspace}
\allsectionsfont{\raggedright}
\renewcommand{\footnotesize}{\fontsize{10}{12}\selectfont}
\AddToHook{cmd/@footnotetext/before}{\def\footnotesize{\setstretch{1}\fontsize{11}{13.2}\selectfont}}
\AddToHook{cmd/@footnotetext/after}{\let\footnotesize\layoutfootnotesize}
\AddToHook{env/figure/begin}{\setstretch{1}\let\small\footnotesize}
\AddToHook{env/table/begin}{\setstretch{1}\let\small\footnotesize}
\AddToHook{env/abstract/begin}{\singlespacing}
\AddToHook{env/thebibliography/begin}{\setstretch{1.1}}
\definecolor{manuscriptlink}{HTML}{1F4E79}
\hypersetup{colorlinks=true,linkcolor=manuscriptlink,citecolor=manuscriptlink,urlcolor=manuscriptlink,filecolor=manuscriptlink,anchorcolor=manuscriptlink}

\newcommand{\manuscripttitle}[2]{%
  \begin{center}
  \setstretch{1.1}
  {\fontsize{16}{20}\selectfont\bfseries #1\par}%
  \if\relax\detokenize{#2}\relax\else
    \vspace{6pt}%
    {\fontsize{16}{20}\selectfont\bfseries #2\par}%
  \fi
  \end{center}
  \vspace{12pt}
}
\renewcommand{\exhibitnotes}[2][\smallskipamount]{%
  \par\vspace{#1}
  \begin{minipage}{0.96\linewidth}
  \singlespacing\footnotesize\setlength{\parindent}{0pt}%
  \textit{Notes:} #2
  \end{minipage}%
}

\hypersetup{pdftitle={AI Search as an Information Intermediary: Clickstream Evidence on Web Traffic}}

\usepackage{subcaption,rotating,fancyvrb,fvextra,tablefootnote,placeins}
\hypersetup{pdfauthor={Qiaoni Shi, Kai Gu, Kai Zhu}}
\makeatletter
\g@addto@macro\UrlBreaks{\do\.\do\-\do\_\do\/}
\makeatother
\definecolor{okBlue}{HTML}{0072B2}
\definecolor{okOrange}{HTML}{E69F00}
\definecolor{okSky}{HTML}{56B4E9}
\definecolor{okGreen}{HTML}{009E73}
\definecolor{okYellow}{HTML}{F0E442}
\definecolor{okVerm}{HTML}{D55E00}
\definecolor{okPink}{HTML}{CC79A7}

\begin{document}

\pagenumbering{arabic}

\manuscripttitle{AI Search as an Information Intermediary:}{Clickstream Evidence on Web Traffic}
\begin{center}
{\large Qiaoni Shi, Kai Gu, and Kai Zhu}\\[0.6em]
{\normalsize Bocconi University, Milan, Italy}\\[0.4em]
{\normalsize This version: November 12, 2026}
\end{center}

\begin{abstract}
\setstretch{1.15}
\noindent AI search retrieves web information to ground generated answers. We study this new information intermediary using U.S. desktop clickstream data observing use alongside referrals during ChatGPT Search's rollout. Referrals accompany 5.4\% of ChatGPT sessions versus 31.1\% of Google queries. Ad-supported sites receive a smaller share of ChatGPT than Google referrals, while reference and technical sites receive larger shares. Conventional searches (Google, Bing, Yahoo) decline after ChatGPT adoption, sooner when adoption follows ChatGPT Search's expansion to users without accounts. Among existing users, that expansion raises ChatGPT use; conventional searches show no detectable overall change but fall among households with lower prior search activity. These findings suggest AI search can reshape open-web traffic through its own referrals and changes in conventional search.
\par\bigskip
\noindent\textbf{Keywords:} AI search; information intermediation; referral traffic; clickstream data; search engines.
\end{abstract}

\clearpage
\section{Introduction}\label{sec:intro}

In November~2026, OpenAI reported that ChatGPT's search volume had reached one-fifth of Google's, counting only requests that trigger a web search.\footnote{Hayden Field, ``OpenAI says ChatGPT is catching up to Google in search,'' \emph{The Verge}, November~3, 2026, \url{https://www.theverge.com/ai-artificial-intelligence/1025482/openai-says-chatgpt-is-catching-up-to-google-in-search} (accessed November~4, 2026). OpenAI does not count requests answered without a web search, such as defining a term, and the report does not state the source of the Google figure.} Regulation also recognizes ChatGPT as a search service: in August~2026, the European Commission designated it a ``very large online search engine'' under the Digital Services Act.\footnote{European Commission, ``Commission designates ChatGPT, Reddit, Roblox under Digital Services Act,'' August~31, 2026, \url{https://digital-strategy.ec.europa.eu/en/news/commission-designates-chatgpt-reddit-roblox-under-digital-services-act} (accessed September~21, 2026).} Where people search matters for websites. Search engines are information intermediaries: they help users locate information and direct traffic to external sources \citep{bakos1997}; for websites, these visits connect information provision with opportunities to attract readers, customers, and contributors \citep{dellarocas2013}. AI search introduces a new form of intermediation, raising the question of how that connection changes when users receive answers within the intermediary.

\emph{AI search} refers here to a system that retrieves web information and uses a large language model to synthesize an answer grounded in it (\citealp{nakano2021}; \autoref{fig:routing-schema}). Web retrieval adds a capability to language generation: it lets the system consult external sources at the time of a request, so its answer can incorporate newly published or updated information beyond the model's training knowledge. OpenAI introduced ChatGPT Search on October~31, 2024, and subsequently expanded access to additional user groups \citep{openai2024search}.
\begin{figure}[H]
\caption{AI search can use a web source without sending the user to it.}
\label{fig:routing-schema}
\centering
\begingroup
\begin{tikzpicture}[
  >={Stealth[length=2.1mm,width=1.5mm]},
  actor/.style={rounded corners=3pt, line width=0.7pt,
    align=center, inner sep=9pt, minimum height=2.25cm},
  flow/.style={->, line width=0.9pt, draw=black},
  request/.style={->, line width=0.7pt, draw=black},
  flowlabel/.style={font=\footnotesize, align=center, inner sep=3pt},
  every node/.style={font=\footnotesize}
]
\node[actor, draw=black, fill=black!4, text width=2.2cm]
  (user) at (0,0)
  {{\large\bfseries User}\\[5pt]Information\\need};
\node[actor, draw=black, fill=black!4, text width=2.9cm]
  (ai) at (5.2,0)
  {{\large\bfseries\color{black}AI search}\\[5pt]Retrieve sources\\Generate answer};
\node[actor, draw=black, fill=black!4, text width=3.0cm]
  (web) at (10.75,0)
  {{\large\bfseries\color{black}Web sources}\\[5pt]New and updated\\information};
\draw[request] ([yshift=0.32cm]user.east) --
  node[flowlabel, above] {Query} ([yshift=0.32cm]ai.west);
\draw[flow] ([yshift=-0.32cm]ai.west) --
  node[flowlabel, below, text=black] {Grounded\\answer}
  ([yshift=-0.32cm]user.east);
\draw[request] ([yshift=0.32cm]ai.east) --
  node[flowlabel, above] {Retrieve} ([yshift=0.32cm]web.west);
\draw[flow] ([yshift=-0.32cm]web.west) --
  node[flowlabel, below, text=black] {Retrieved\\material}
  ([yshift=-0.32cm]ai.east);
\draw[->, draw=black, line width=1pt,
  dash pattern=on 4pt off 3pt, rounded corners=6pt]
  (user.south) -- (0,-2.0) --
  node[flowlabel, below=3pt, text=black]
    {Optional user visit to a source}
  (10.75,-2.0) -- (web.south);
\end{tikzpicture}
\endgroup

\exhibitnotes[-6pt]{Solid arrows show the user's query, the system's retrieval of web information, and the generated answer. The dashed arrow shows an optional user visit to a source---a referral. Only some ChatGPT sessions involve web retrieval.}
\end{figure}

For websites, supplying information to AI search need not generate a visit: users can read the answer without opening its sources. Yet AI search can also help users discover sources and encourage visits---a tension familiar from research on content aggregation \citep{dellarocas2016,chiou2017,calzada2020}. Its effects on websites run through two channels: first, its own referrals; second, changes in conventional search (Google, Bing, and Yahoo) and the visits it generates. Measuring both takes more than a count of AI searches: it requires following users to the websites they visit and to the conventional searches they make.

We study both channels using URL-level Comscore U.S. desktop clickstream data from October~2024 through July~2025, spanning the initial ChatGPT Search rollout. These records follow households across the web, but do not directly identify conversations, queries, or referrals: a single ChatGPT conversation can produce many page loads and background requests, and the source of a website visit must be inferred from its referrer and timing. We reconstruct ChatGPT sessions, Google queries, and attributed referrals to external websites, and follow households' weekly conventional searches in a balanced panel. Observing intermediary activity alongside external referrals lets us count occasions that generate no recorded referral. These belong in the denominator of referral incidence: the share of ChatGPT sessions or Google queries associated with at least one referral. This distinguishes how much an intermediary is used from how often that use connects users with external websites.

For the first channel, we find substantially lower referral incidence for ChatGPT than for Google Search, the established intermediary we use as a benchmark. A referral to an external website accompanies 5.4\% of ChatGPT sessions; the corresponding figure for Google queries is 31.1\%. Over the ten months, 74.4\% of households that use ChatGPT have no recorded referral from it, compared with 9.6\% of households that query Google. The difference persists within households and after adjustment for the types of websites households browse around their use of each intermediary. The comparison describes the intermediaries as used: they serve different tasks in different activity units, and some ChatGPT conversations are answered without web retrieval. This is the referral channel websites encounter in practice.

The referrals ChatGPT does generate reach a distinctive set of destinations. Reference, tools/SaaS, academic, and developer websites account for larger shares of ChatGPT referrals than of Google referrals, while ad-supported websites account for a smaller share. When pooled across households, ChatGPT referrals are less concentrated across destinations than Google's. The household comparisons show no corresponding reduction in concentration, and ChatGPT referrals are more concentrated among household-weeks with more referrals. This pattern is consistent with different households directing their referrals to different websites.

For the second channel, we ask how households' conventional searches change once they begin using ChatGPT. We use a balanced panel of \panelN{} households and two complementary designs. As in \citet{padilla2025}, the first design compares households that adopt ChatGPT with households that adopt later. Weekly Google, Bing, and Yahoo searches rise briefly at adoption and then fall by about 39\% of the pre-adoption average in weeks 20--30. This decline need not come from AI search, since ChatGPT can answer many questions without searching the web. The rollout offers a comparison: households adopting on or after February~5, 2025 could use Search from their first week without an account, whereas earlier adopters began under more restricted access. At the same number of weeks since adoption, conventional searches decline sooner among later adopters, providing suggestive evidence of a role for ChatGPT Search beyond general LLM adoption.

Because households choose when to adopt, the second design uses the February~5 expansion itself, a date set by OpenAI \citep{openai2024search}. Households already using ChatGPT without an account gained Search on that date; the design compares them with prior Gemini users, who had likewise used an AI service before the expansion. Newly eligible households record more ChatGPT sessions afterward, without a detectable overall change in conventional search. The search response differs significantly with prior search activity, which reflects how heavily households rely on conventional search and possibly what they use it for. Households with lower prior activity reduce weekly conventional searches by 3.71, or 35.8\% of their pre-expansion mean, whereas the estimate for higher-activity households is positive and imprecise. ChatGPT sessions rise by similar amounts in both groups, so the difference lies in how conventional search responds.

Differences in what households search for offer a possible explanation. Before the expansion, lower-activity households direct larger shares of their conventional-search referrals to tools/SaaS and e-commerce brand websites, and higher-activity households larger shares to social media and adult websites. After the expansion, such referrals to tools/SaaS and brand websites decline in the full access sample, with estimates significant at the 10\% level. These patterns are consistent with substitution for some task-oriented and product-related searches, while searches for content or interaction on a platform itself may be harder to replace.

Our first contribution is to measure and characterize the referral role of AI search, a new information intermediary. Existing work examines how generated summaries within Google Search affect traffic to publishers and Wikipedia \citep{agarwal2026,khosravi2026,wang2026}, the commercial outcomes of arriving ChatGPT referrals \citep{kaiser2026}, and how LLM adoption changes the concentration of web traffic \citep{gholami2026beyond}. We complement it by reconstructing ChatGPT sessions, search queries, and their referrals from clickstream records, which allows a direct comparison of ChatGPT with Google under a common referral definition during the rollout of ChatGPT Search: how often use is accompanied by a referral, which websites receive those referrals, and how concentrated they are across the web and within households.

Second, we examine how conventional search changes when households gain access to ChatGPT's Search feature. This extends evidence on LLM adoption by examining the role of web search within ChatGPT. \citet{padilla2025} document search declines after households adopt LLMs in 2022--2023, before the launch of ChatGPT Search, whereas \citet{gholami2026beyond} find a short-lived rise in search and suggest that LLMs can complement it. In our data, both patterns appear in the same households at different horizons: conventional search rises in the first weeks after adoption and declines later. The access expansion, whose timing households did not choose, adds evidence that gaining Search reduces conventional search among households with lower prior search activity, with no detectable change overall. Substitution and complementarity may therefore both hold, depending on the horizon and the household, and possibly on the needs that search serves.

\autoref{sec:framework} develops the conceptual framework. \autoref{sec:setting} describes the data and methods, \autoref{sec:results} presents the referral findings, \autoref{sec:displacement} examines how ChatGPT adoption and expanded access to ChatGPT Search change conventional search, and \autoref{sec:conclusion} discusses their implications.

\section{Conceptual Framework}\label{sec:framework}

AI search can affect website traffic through two channels. The first consists of referrals generated by AI search itself. The second operates through changes in conventional search and the referrals it produces. Through the first channel, AI search can change both the need for an external visit and the reason for making one. An answer that supplies the information a user seeks may substitute for a visit, while one that reveals a useful source may encourage further exploration. Research on content aggregation documents this tension between supplying information within an intermediary and helping users discover external content \citep{dellarocas2016,chiou2017,calzada2020}.

For AI search, the remaining value of a website visit may lie in examining detailed evidence, using an interactive tool, or completing a transaction. These purposes connect referral incidence with destination composition: visits may decline for needs addressed within the answer while continuing for activities that require a website. The destinations receiving those visits can also differ from the sources used to construct the answer. Referral patterns therefore reveal which websites users continue to visit alongside AI search.

The distribution of those referrals depends on both individual choices and their overlap across households. Lower search costs can help users find niche offerings \citep{brynjolfsson2011}, but broader discovery by individuals can coexist with greater aggregate concentration when their choices converge on the same destinations \citep{fleder2009}.

The reverse is also possible. Households could each concentrate their referrals on a few specialist websites yet collectively generate a dispersed referral stream because they visit different sites. Aggregate dispersion can thus arise from differences across households as well as from dispersion within them. Examining both levels distinguishes how strongly an intermediary's traffic converges on particular websites from how evenly individual households distribute their referrals.

The second channel runs through conventional search. An answer may replace a conventional-search query, but it may also introduce a topic or source that prompts further searching. Web retrieval widens the questions an answer can address. A model answering from its training knowledge can help with a calculation or a draft, but questions about current products, prices, or events would otherwise send the user to a search engine. Access to AI search can therefore affect a wider range of conventional searches than general LLM use.

How much conventional search an answer can replace also depends on what the search is for, a distinction long used to classify web queries \citep{broder2002}. Some searches seek information or help that an answer can supply directly, such as a calculation, a piece of writing, or product information. Others lead to content or interaction available only on a particular website, such as a video or a discussion. Households can differ in this mix of needs, so the same access can reduce search for some and leave it unchanged for others; an average response can conceal the difference. The destinations a household's searches lead to offer an observable indication of the mix. Because only some queries lead to external visits, the effect on search-referred visits depends on which searches change.

For a destination website, the two channels combine. When AI search replaces a conventional search, the visit that search would have produced depends on the user continuing from the answer to the website. A website could therefore account for a relatively large share of AI referrals yet receive fewer visits overall. How much traffic it retains turns on the distinction running through both channels: whether an answer can meet the user's need, or only the website can.

\section{Data and Method}\label{sec:setting}

\subsection{Data coverage and observable signals}\label{sec:data-coverage}

We use URL-level records from the Comscore U.S. Desktop Panel from October~2024 through July~2025. The records follow the same households across ChatGPT, conventional search engines, and destination websites during the introduction and expansion of ChatGPT Search (\autoref{fig:timeline}). Earlier evidence on LLM adoption covers 2022--2023, before the introduction of this feature \citep{padilla2025}. Our observation period captures households gaining access to Search, including households already using ChatGPT. This coverage allows us to characterize ChatGPT's referrals during the rollout and estimate how gaining its Search feature changes conventional search and the visits it generates.

\begin{figure}[!htbp]
\caption{The observation window spans the three ChatGPT Search access expansions.}
\label{fig:timeline}
\centering
\begingroup
\definecolor{expansionRed}{HTML}{C00000}
\begin{tikzpicture}[
  x=1cm, y=1cm,
  every node/.style={font=\footnotesize},
  lab/.style={font=\footnotesize, align=center, inner sep=1.5pt},
  redlab/.style={lab, text=expansionRed},
  graylab/.style={lab, text=black!60},
  axis/.style={line width=0.8pt, draw=black},
  tick/.style={line width=0.6pt, draw=black},
  stem/.style={line width=0.7pt, draw=expansionRed}
]
\draw[axis] (0.3,0) -- (1.55,0);
\draw[axis] (1.55,0) -- (1.68,0.14) -- (1.84,-0.14) -- (1.97,0);
\draw[axis] (1.97,0) -- (15.5,0);
\fill[black!60] (1.0,0) circle (2pt);
\node[graylab, anchor=south, text width=2.4cm] at (1.0,0.3) {Nov 30, 2022\\ChatGPT released};
\node[graylab, anchor=north] at (1.0,-0.12) {Dec 2022};
\foreach \k in {0,...,10} { \draw[tick] (2.6+\k*1.26,0.1) -- (2.6+\k*1.26,-0.1); }
\foreach \k/\m in {0/{Oct 2024},1/Nov,2/Dec,3/{Jan 2025},4/Feb,5/Mar,6/Apr,7/May,8/Jun,9/Jul} {
  \node[lab, anchor=north] at (2.6+\k*1.26+0.63,-0.12) {\m}; }
\draw[tick] (2.6,-0.7) -- (2.6,-0.85) -- (15.2,-0.85) -- (15.2,-0.7);
\node[lab, anchor=north] at (8.9,-0.88) {Comscore U.S. desktop panel, October 2024 -- July 2025};
\draw[stem] (3.82,0) -- (3.82,1.5);
\fill[expansionRed] (3.82,0) circle (2.4pt);
\node[redlab, anchor=south, text width=2.6cm] at (3.82,1.5) {\textbf{Oct 31, 2024}\\Paid subscribers};
\draw[stem] (5.73,0) -- (5.73,0.5);
\fill[expansionRed] (5.73,0) circle (2.4pt);
\node[redlab, anchor=south, text width=3.5cm] at (5.73,0.5) {\textbf{Dec 16, 2024}\\Free logged-in users};
\draw[stem] (7.82,0) -- (7.82,1.5);
\fill[expansionRed] (7.82,0) circle (2.4pt);
\node[redlab, anchor=south, text width=2.6cm] at (7.82,1.5) {\textbf{Feb 5, 2025}\\Users without\\an account};
\end{tikzpicture}
\endgroup

\exhibitnotes[-6pt]{Red markers show the introduction of ChatGPT Search for paid subscribers and its subsequent expansion to free logged-in users and users without an account \citep{openai2024search}. Our Comscore U.S. desktop data cover October~2024 through July~2025, spanning all three access expansions. The break in the axis separates ChatGPT's November~2022 release from our observation window.}
\end{figure}

The records combine URLs and timestamps with source referrers and request characteristics. The full sample contains between 168{,}467 and 238{,}315 active households per month. For the analysis of conventional search, we use a balanced panel of \panelN{} households with at least four webpage visits in every month.\footnote{Sections and tables numbered OA refer to the Online Appendix. See \appref{oa:panel-threshold} for the panel threshold.}

A webpage or conversation can generate many network requests, so measuring use requires separating user actions from background requests. We count a webpage visit only when the record is a successfully loaded HTML page, and we use specific ChatGPT requests to reconstruct conversations and whether a user was signed in.\footnote{See \apprefrange{oa:cleaning}{oa:endpoint-coverage} for filtering rules and the ChatGPT requests used.}

\subsection{Activity and referral measures}\label{sec:activity-referrals}

For referral incidence, we measure conventional-search activity as queries and ChatGPT activity as sessions. Queries are identified from foreground Google, Bing, and Yahoo URLs matching their search patterns. ChatGPT runs as a single-page application, so one conversation can produce many page loads. We treat loads of the same conversation on the same computer as one ChatGPT session, starting a new session when more than an hour passes between loads; different conversations remain separate sessions.\footnote{See \apprefrange{oa:endpoint-coverage}{oa:conv-cleaning} for request paths and sensitivity to alternative session boundaries.} The resulting measure provides an observed denominator for assessing how frequently ChatGPT use generates visits to external websites.

To associate external visits with intermediary use, we combine source referrers with timestamps. A clean referral is a foreground visit to an external website whose HTTP referrer identifies the source intermediary, after excluding visits within the platform itself, sign-in pages, technical infrastructure, and search services. Timestamps link these visits to observed queries or ChatGPT sessions; visits without an observable source referrer remain unattributed.\footnote{See \appref{oa:referral-def} for attribution windows, exclusions, and validation checks.}

Referral incidence is the share of ChatGPT sessions or queries associated with at least one clean referral. Each occasion enters the numerator at most once, while occasions without a referral remain in the denominator; their share is absorptiveness.

\subsection{Destinations and surrounding context}\label{sec:destination-measures}

An intermediary's relationship with the web also depends on which websites receive its referrals. We compare ChatGPT and Google in the composition of their referral destinations, examining both the content websites provide and how they monetize their activity. We classify 4{,}266 destination domains by content type and monetization model using GPT-4o and a common codebook. Content-type analyses use the 3{,}245 domains with classifier confidence of at least 0.90.\footnote{See \appref{oa:classification} for classification validation, including agreement rates and category-level disagreements, and sensitivity to the confidence restriction.}

A further question is whether referrals converge on a few websites or spread across many. We measure concentration using a floor-corrected normalized Herfindahl--Hirschman index (HHI), calculated separately across aggregate referrals and within household-weeks. The former captures whether households' referrals converge on the same destinations; the latter captures how broadly each household's referrals are distributed. This distinction allows aggregate dispersion to coexist with concentrated referrals within individual households (\appref{oa:concentration}).

Interpreting referral differences also requires characterizing the contexts in which the intermediaries are used. Surrounding context refers to the other websites a household visits in the fifteen minutes before and after a ChatGPT session or Google search session. Because prompt text is unavailable, we use the content categories of these websites to characterize the circumstances of intermediary use. A category is dominant when it accounts for at least half of this browsing. Sessions with no classifiable surrounding browsing are labeled solo; those with browsing but no dominant category are labeled mixed and excluded from the category comparison (\appref{oa:context}). We compare the intermediaries' surrounding-context shares and measure ChatGPT's referral incidence within each context, distinguishing where use is common from where it most frequently generates external visits.

\subsection{ChatGPT Search access and conventional-search measures}\label{sec:access-comparisons}

To examine changes in conventional search, we follow the \panelN{} households in the balanced panel before and after they adopt ChatGPT or gain access to Search.\footnote{Both comparisons use ACS age-by-income weights to align their samples with the U.S. population; \appref{oa:acs-reweighting} describes the weighting procedure.} Weekly conventional searches sum the Google, Bing, and Yahoo query records defined in \autoref{sec:activity-referrals}. Each recorded result-page load contributes one count, including repeated loads of the same query. We measure ChatGPT use using the same session definition.

\begin{figure}[!htbp]
\centering
\captionsetup{skip=2pt}
\caption{ChatGPT reach rises during the rollout of ChatGPT Search while Google reach remains stable.}
\label{fig:adoption}
\begin{minipage}[t]{0.48\linewidth}
\centering
\includegraphics[width=\linewidth]{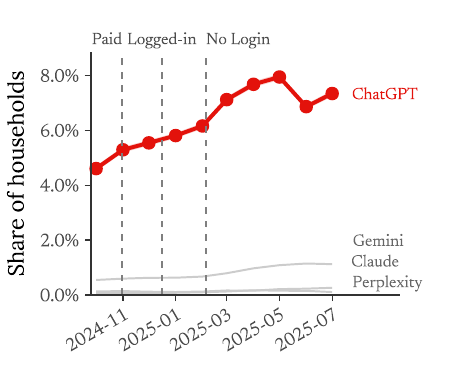}
\smallskip\small (a) Households visiting an LLM platform.
\end{minipage}
\hfill
\begin{minipage}[t]{0.48\linewidth}
\centering
\includegraphics[width=\linewidth]{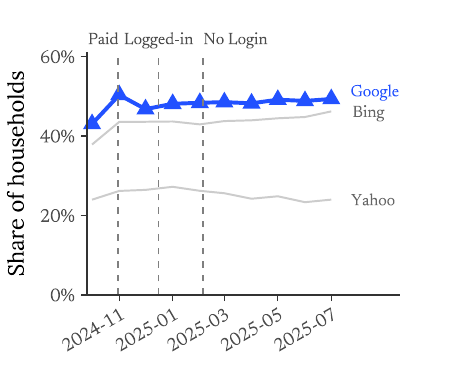}
\smallskip\small (b) Households issuing a search query.
\end{minipage}
\exhibitnotes[-6pt]{Each series reports the monthly share of active Comscore U.S. desktop households that visits an LLM platform (panel a) or issues a search-engine query (panel b). Dashed vertical lines mark the October~31, December~16, and February~5 ChatGPT Search access expansions. ChatGPT reach rises from 4.6\% in October~2024 to a peak near 7.9\%; Google remains near 49\%.}
\end{figure}

\autoref{fig:adoption} shows that ChatGPT reaches a growing share of active households during the Search rollout, while Google's reach remains broadly stable. Our household-level comparisons examine how the frequency of conventional search changes as households adopt ChatGPT or gain access to Search.

\autoref{sec:displacement} uses two complementary comparisons. The first follows households from the beginning of sustained ChatGPT use, using households that have yet to adopt as comparisons. The second compares households that previously used ChatGPT without an account with prior Gemini users around the February expansion, estimating how gaining Search changes conventional search among existing ChatGPT users.

\section{ChatGPT as a Referral Intermediary}\label{sec:results}

AI search affects web traffic directly through its referrals to external websites. This first channel raises three questions. How often does use lead to an external visit at all? When it does, which websites, and which business models, receive the visits? And do those visits converge on a few destinations or spread across many?

\subsection{ChatGPT sessions have low referral incidence}\label{sec:absorption}

To estimate how frequently ChatGPT use leads to an external visit, we calculate the share of ChatGPT sessions associated with at least one clean outbound referral. Sessions without a recorded referral remain in the denominator. We compare this referral incidence with the corresponding share of Google Search queries.

\autoref{tab:absorption-summary} shows that a clean outbound referral accompanies 5.4\% of ChatGPT sessions, compared with 31.1\% of Google Search queries. The absence of recorded referrals is also widespread across households: over the ten-month observation period, 74.4\% of ChatGPT-active households generate no recorded clean ChatGPT referral, compared with 9.6\% of Google-querying households.\footnote{Nonreferral, or absorptiveness, can reflect a satisfactory answer, an abandoned task, or an external visit with a missing referrer. See \appref{oa:ctr-bench} for comparisons with selected public benchmarks, accounting for differences in denominators, attribution windows, population coverage, and traffic filters.}

\begin{table}[!htbp]
\centering
\caption{ChatGPT sessions have lower referral incidence than Google queries.}
\label{tab:absorption-summary}
\begin{tabular}{lrr}
\toprule
 & Google & ChatGPT \\
\midrule
Occasion of use & Search query & Session \\
Occasions & 61{,}689{,}308 & 409{,}133 \\
Clean referral visits & 31{,}464{,}344 & 168{,}111 \\
Referral incidence & 31.1\% & 5.4\% \\
Active households & 291{,}747 & 56{,}578 \\
Households with no clean referral & 9.6\% & 74.4\% \\
\bottomrule
\end{tabular}
\par\vspace{-12pt}
\exhibitnotes{The sample covers Comscore U.S. desktop activity from October~2024 through July~2025. An occasion of use is one Google query or one ChatGPT session. Clean referral visits are visits to third-party websites that meet the referral definition in \autoref{sec:activity-referrals}; referral incidence is the share of occasions producing at least one such visit. The last row is the share of active households with no clean referral from that intermediary during the ten months.}
\end{table}

\begin{figure}[!htbp]
\centering
\captionsetup{skip=2pt}
\caption{Referral incidence is lower for ChatGPT.}
\label{fig:absorption-ratio}
\begin{minipage}[t]{0.48\linewidth}
\centering
\includegraphics[width=\linewidth]{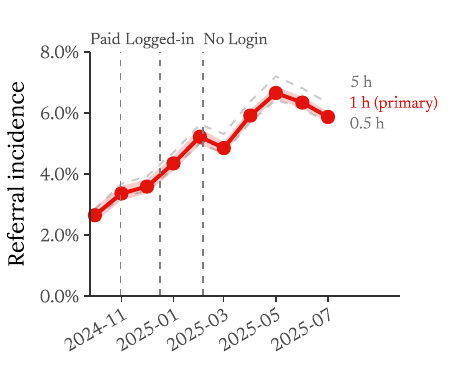}
\smallskip\small (a) ChatGPT sessions.
\end{minipage}
\hfill
\begin{minipage}[t]{0.48\linewidth}
\centering
\includegraphics[width=\linewidth]{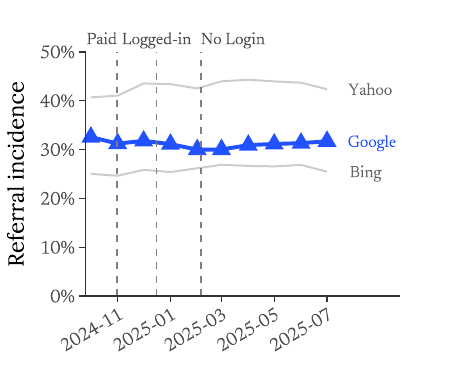}
\smallskip\small (b) Search-engine queries.
\end{minipage}
\exhibitnotes[-6pt]{Points show the monthly share of ChatGPT sessions (panel a) or search queries (panel b) associated with at least one clean referral. In panel (a), the red series uses a one-hour ChatGPT session boundary; gray series use thirty-minute and five-hour boundaries. Panel (b) shows Google, Bing, and Yahoo queries, with Google highlighted. Dashed lines mark the three ChatGPT Search access expansions.}
\end{figure}

The difference persists throughout the observation window: even though ChatGPT's monthly referral incidence rises from approximately 2.6\% to a peak near 6.7\% (\autoref{fig:absorption-ratio}), it remains far below Google's per-query incidence. As shown in \autoref{fig:absorption-ratio}, we also vary the ChatGPT session boundary from thirty minutes to five hours; ChatGPT's referral incidence remains low.

\paragraph{Accounting for household and week differences.} \autoref{tab:absorption-summary} compares aggregate rates, which pool all households and weeks. The lower ChatGPT incidence could thus reflect a few heavy users or particular periods rather than a difference for a given household in a given week. We therefore aggregate referral incidence by household, week, and intermediary and estimate
\begin{equation}
\text{Referral incidence}_{itq}
= \beta\,\mathbf{1}\{q=\text{ChatGPT}\}+\alpha_i+\gamma_t+\varepsilon_{itq},
\label{eq:ref-ratio}
\end{equation}
where the outcome is intermediary $q$'s referral incidence for household $i$ in week $t$. Household fixed effects account for persistent differences across households, and week fixed effects account for common calendar changes. Standard errors are clustered by household. \autoref{tab:referral-comparisons} shows that lower ChatGPT referral incidence persists after adjustment for household and week effects: the estimated difference is 29.0 percentage points ($\hat\beta=-0.290$, $p<0.001$).

\paragraph{Accounting for surrounding context.} Household and week fixed effects hold the household fixed but not the task. The same household may bring different purposes to ChatGPT and to Google, and purposes differ in how often they require a visit. Prompts and queries are unobserved, so we use surrounding browsing as a proxy for the task: the content category of the other websites a household visits in the fifteen minutes around a session (\autoref{sec:destination-measures}), such as reference and knowledge, tools and SaaS, or social media. We label each ChatGPT session by its surrounding browsing and each Google query by the context of the search session it belongs to, compute referral incidence separately for each context within each household, week, and intermediary, and add context fixed effects, so that sessions are compared within the same context. Because this sample excludes sessions without a classifiable context, it differs from that of columns (1)--(3).\footnote{Columns (4) and (5) nevertheless have more observations, because incidence is computed separately for each context: a household's activity through one intermediary in one week contributes one observation in columns (1)--(3) and one per context in columns (4) and (5).} Adding context fixed effects changes the ChatGPT coefficient only slightly, from $-0.334$ to $-0.329$ (\autoref{tab:referral-comparisons}; both $p<0.001$). ChatGPT's lower referral incidence thus persists when sessions in the same context are compared. \autoref{sec:contexts} describes how the contexts of ChatGPT and Google use differ.

\begin{table}[!htbp]
\centering
\caption{ChatGPT has a lower referral incidence.}
\label{tab:referral-comparisons}
\small
\begin{tabular}{lccccc}
\toprule
 & \multicolumn{5}{c}{Dependent variable: referral incidence} \\
\cmidrule(lr){2-4}\cmidrule(lr){5-6}
 & (1) & (2) & (3) & (4) & (5) \\
\midrule
ChatGPT indicator & $-0.315^{***}$ & $-0.289^{***}$ & $-0.290^{***}$ & $-0.334^{***}$ & $-0.329^{***}$ \\
 & (0.001) & (0.002) & (0.002) & (0.002) & (0.002) \\
\addlinespace
Observations split by context & No & No & No & Yes & Yes \\
Household fixed effects & No & Yes & Yes & Yes & Yes \\
Week fixed effects & No & No & Yes & Yes & Yes \\
Context fixed effects & No & No & No & No & Yes \\
$R^2$ & 0.034 & 0.367 & 0.368 & 0.235 & 0.244 \\
Observations & 2{,}790{,}136 & 2{,}790{,}136 & 2{,}790{,}136 & 3{,}809{,}132 & 3{,}809{,}132 \\
\bottomrule
\end{tabular}
\par\vspace{-12pt}
\exhibitnotes{The sample covers Comscore U.S. desktop activity from October~2024 through July~2025. Referral incidence is measured on a zero-to-one scale; the ChatGPT indicator compares ChatGPT with Google. In columns (1)--(3), each observation represents one household, week, and intermediary. Columns (4)--(5) further split observations by surrounding browsing context and exclude activity without a classifiable context. Observations are equally weighted. Household-clustered standard errors appear in parentheses. \appref{oa:within-user-ref} provides construction and sample details. $^*p<0.10$, $^{**}p<0.05$, $^{***}p<0.01$.}
\end{table}

\subsection{ChatGPT referrals have a different destination mix}\label{sec:selective}

When referrals occur, which parts of the web receive them? The question matters to publishers and other website operators, who want to know whether they can expect visitors from AI search and how its referrals differ from those of conventional search.

To answer it, we classify the destination websites of ChatGPT and Google referrals, the external websites users visit from each intermediary, by content type, using the categories defined in \autoref{sec:destination-measures}, such as reference and knowledge, social media, or e-commerce marketplaces. For each category we compute its share of ChatGPT's classified referrals and its share of Google's, and compare the two.

\begin{figure}[!htbp]
\centering
\caption{Tools and knowledge sites have larger shares of ChatGPT referrals; ad-supported sites have a smaller share.}
\label{fig:composition}
\begin{minipage}[t]{0.495\linewidth}
\centering
\includegraphics[width=\linewidth]{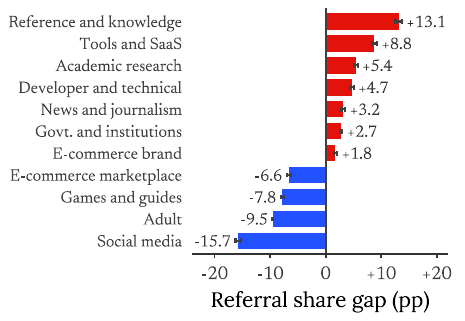}
\smallskip\small (a) ChatGPT minus Google referral share by content type.
\end{minipage}
\hfill
\begin{minipage}[t]{0.495\linewidth}
\centering
\includegraphics[width=\linewidth]{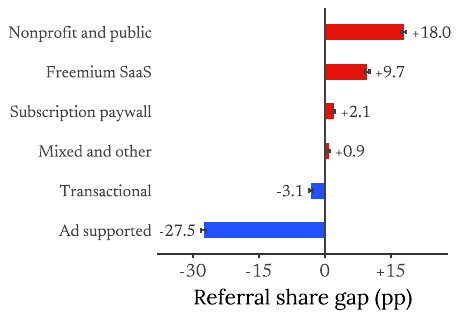}
\smallskip\small (b) ChatGPT minus Google referral share by monetization model.
\end{minipage}
\exhibitnotes[0pt]{The sample includes 3{,}245 destination domains classified with confidence of at least 0.90 (\appref{oa:classification}). Bars and labels show each category's share of ChatGPT referrals minus its share of Google referrals, in percentage points. Error bars show 95\% pointwise normal confidence intervals for differences in proportions.}
\end{figure}

\autoref{fig:composition}, panel~(a), plots the difference in shares; a positive value means the category accounts for a larger share of ChatGPT referrals than of Google referrals. Knowledge and tools account for larger shares of ChatGPT's referrals than of Google's. Reference/knowledge websites account for a 13.1 percentage-point larger share, followed by tools/SaaS at 8.8 points, academic research at 5.4 points, and developer/technical websites at 4.7 points. Social media, by contrast, accounts for a 15.7-point smaller share.

In addition, we classify those referred websites by how they finance their content and services (\autoref{fig:composition}, panel~b). Ad-supported websites account for a 27.5-percentage-point smaller share of ChatGPT referrals than of Google referrals. Non-profit/public, freemium SaaS, and subscription websites account for larger shares, while transactional destinations account for smaller shares. Both patterns persist when the classifier-confidence threshold is relaxed from 0.90 to 0.60 (\appref{oa:composition-confidence}).

\subsection{Concentration in aggregate and within households}\label{sec:concentration}

Do the referrals of AI search converge on a few destinations or spread across many? The answer matters to publishers of every size. For large publishers, it shows whether the position they hold as the main recipients of conventional-search referrals carries over to AI search. For smaller publishers, it shows whether the referrals AI search does send reach websites that receive few through conventional search, or flow to the same large destinations.

\begin{samepage}
We measure concentration using the normalized Herfindahl--Hirschman index (HHI) of referral shares across destination domains:
\begin{equation}
H^{*}=\frac{\sum_{d=1}^{n}s_d^{2}-1/n}{1-1/n},\qquad n\geq2,
\label{eq:normhhi}
\end{equation}
where $s_d$ is a domain's share of an intermediary's referrals and $n$ is the number of observed destination domains.\footnote{We exclude single-destination observations, for which the normalized index is undefined; see \appref{oa:hhi-def}.} Subtracting $1/n$ removes the minimum concentration associated with that destination count. The index equals zero when referrals are distributed equally across domains and approaches one as a single domain dominates. We calculate it both across households and within household-weeks to distinguish concentration in the overall referral stream from concentration in individual households' referrals.
\end{samepage}

\autoref{tab:hhi} reports the index for each intermediary's referrals pooled over all households. Across all destination domains, Google's index is 3.47 times ChatGPT's. Because the full set includes many destinations with few referrals, we also compute the index among the 100, 1{,}000, 2{,}500, and 10{,}000 most-referred domains. Google's index is higher in every set, with a ratio between 1.87 and 3.47 (\appref{oa:hhi}). Pooled across households, ChatGPT referrals are thus less concentrated than Google's.

\begin{figure}[!htbp]
\centering
\caption{Domains with higher relative ChatGPT referral shares tend to have lower referral volume.}
\label{fig:share-ratio}
\includegraphics[width=0.82\linewidth]{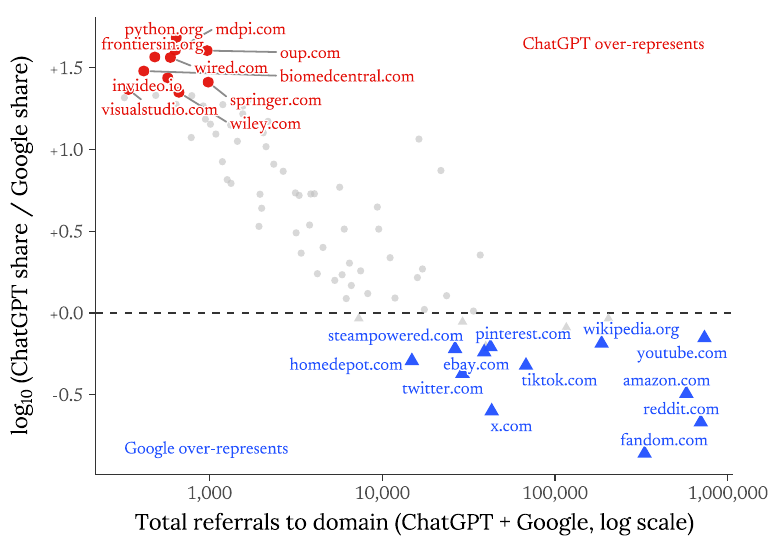}
\exhibitnotes[-6pt]{Each point represents one destination domain. The figure shows the 250 domains with the highest combined ChatGPT and Google referral volume among those reaching at least 0.5\% of each intermediary's referring households. The horizontal axis shows combined referral volume on a logarithmic scale; the vertical axis shows the base-ten logarithm of the ChatGPT-to-Google referral-share ratio. Positive values indicate a larger share of ChatGPT referrals; negative values indicate a larger share of Google referrals.}
\end{figure}

We further examine which destinations produce this aggregate difference. For each destination domain, we plot the logarithm of the domain's share of ChatGPT referrals divided by its share of Google referrals against the domain's total referral volume (\autoref{fig:share-ratio}). A domain with a larger share of ChatGPT referrals lies above the zero line, and one with a larger share of Google referrals lies below it. Large websites such as Reddit, Amazon, and TikTok account for relatively larger shares of Google referrals, while academic publishers, reference websites, and tool-oriented websites account for relatively larger shares of ChatGPT referrals. The destinations that lean toward Google are among the largest in the sample, so Google's referrals converge on a few high-volume websites, whereas ChatGPT's are spread over more and smaller ones.

\begin{table}[!htbp]
\centering
\caption{Google referrals are more concentrated than ChatGPT referrals in aggregate.}
\label{tab:hhi}
\begin{tabular}{lrrr}
\toprule
Destination set & ChatGPT HHI & Google HHI & Google/ChatGPT \\
\midrule
Top 100 & 0.0360 & 0.0674 & 1.87 \\
Top 1{,}000 & 0.0148 & 0.0396 & 2.67 \\
Top 2{,}500 & 0.0113 & 0.0332 & 2.94 \\
Top 10{,}000 & 0.0078 & 0.0270 & 3.47 \\
All domains & 0.0055 & 0.0191 & 3.47 \\
\bottomrule
\end{tabular}
\par\vspace{-12pt}
\exhibitnotes{HHI is the normalized Herfindahl--Hirschman index calculated from each intermediary's aggregate referral shares; higher values indicate greater concentration. Each ``Top'' row includes the union of ChatGPT's and Google's most-referred domains at the stated cutoff, with shares recalculated within that set. The final column divides Google's HHI by ChatGPT's. \appref{tab:oa-hhi} reports additional destination restrictions.}
\end{table}

A less concentrated pooled stream does not imply that each household spreads its referrals widely, since the aggregate dispersion could also arise because different households visit different websites. To understand which is the case, we compute the index for each household, week, and intermediary, and compare ChatGPT with Google within the same household and week, with household and week fixed effects. Because the index is coarse for a household-week with few referrals, we repeat the comparison among household-weeks with at least one, two, five, and ten referrals from each intermediary. \autoref{tab:within-user-hhi} reports the results. At the one- and two-referral thresholds, the ChatGPT--Google difference is indistinguishable from zero. At the five- and ten-referral thresholds, the ChatGPT coefficient is positive, at 0.035 and 0.056, respectively, so among more active household-weeks ChatGPT referrals are more concentrated than Google's. We thus find no evidence of lower concentration through ChatGPT within households.\footnote{See \appref{oa:concentration} for the full threshold comparisons and supplementary destination characteristics.}

\begin{table}[!htbp]
\centering
\caption{ChatGPT has higher within-household concentration at the five- and ten-referral thresholds.}
\label{tab:within-user-hhi}
\begin{tabular}{l*{4}{>{\centering\arraybackslash}p{0.105\linewidth}}}
\toprule
 & \multicolumn{4}{c}{Dependent variable: normalized HHI} \\
\cmidrule(lr){2-5}
 & \multicolumn{4}{c}{Minimum referrals from each intermediary} \\
 & $\geq 1$ & $\geq 2$ & $\geq 5$ & $\geq 10$ \\
\midrule
ChatGPT indicator & $-0.001$ & $0.000$ & $0.035^{***}$ & $0.056^{***}$ \\
                  & (0.002)  & (0.002) & (0.003)       & (0.006)       \\
\addlinespace
Mean normalized HHI, Google  & 0.072 & 0.071 & 0.075 & 0.077 \\
Mean normalized HHI, ChatGPT & 0.069 & 0.070 & 0.107 & 0.130 \\
\addlinespace
Household fixed effects & Yes      & Yes     & Yes           & Yes           \\
Week fixed effects      & Yes      & Yes     & Yes           & Yes           \\
Observations      & 48{,}641 & 35{,}644 & 12{,}684      & 3{,}721       \\
\bottomrule
\end{tabular}
\par\vspace{-12pt}
\exhibitnotes{Normalized HHI measures concentration across each household's referral destinations within a week by intermediary; higher values indicate greater concentration. Columns retain household-weeks meeting the indicated minimum referral count for each intermediary. Observations with fewer than two distinct destinations are excluded. Household-clustered standard errors appear in parentheses. \appref{tab:oa-within-user-conc} reports thresholds from 1 to 20 referrals. $^*p<0.10$, $^{**}p<0.05$, $^{***}p<0.01$.}
\end{table}

Taken together, ChatGPT referrals are less concentrated than Google's in aggregate, in line with \citet{gholami2026beyond}, who find that traffic LLMs route to external websites is no more concentrated than traffic from traditional search. Within households, we detect no difference at the one- and two-referral thresholds, while ChatGPT referrals are more concentrated at the five- and ten-referral thresholds. Examining both levels distinguishes how referrals are distributed across the web as a whole from how individual households distribute their referrals. The contrast is consistent with households concentrating their referrals on different websites, producing a more dispersed referral stream when pooled together.

\subsection{Surrounding contexts of ChatGPT and Google use}\label{sec:contexts}

When households use ChatGPT, what kind of tasks are they doing? We attempt to answer this question because the answer may shed light on what drives the low referral incidence of ChatGPT use. Prompts and queries are unobserved, so we characterize the task indirectly. We classify the other websites a household visits in the fifteen minutes before and after a ChatGPT session or a Google search session into the content categories of \autoref{sec:destination-measures}, and call the dominant category the session's surrounding context, since it describes the browsing within which the session takes place. Sessions with no classifiable browsing nearby are solo. For each context we compute its share of ChatGPT sessions and its share of Google search sessions and compare the two.

\begin{figure}[!htbp]
\centering
\caption{Context prevalence and ChatGPT referral incidence follow different orderings.}
\label{fig:context}
\begin{minipage}[t]{0.495\linewidth}
\centering
\includegraphics[width=\linewidth]{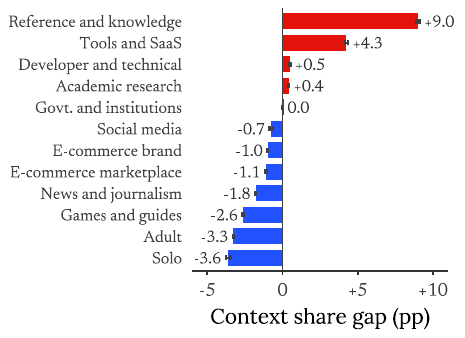}
\smallskip\small (a) ChatGPT minus Google surrounding-context share.
\end{minipage}
\hfill
\begin{minipage}[t]{0.495\linewidth}
\centering
\includegraphics[width=\linewidth]{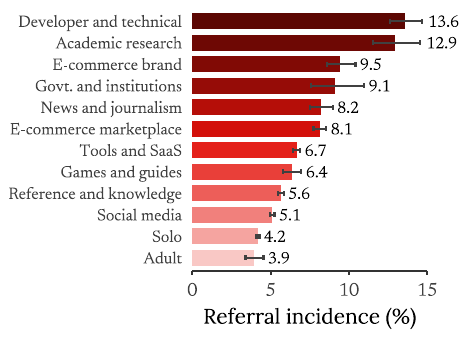}
\smallskip\small (b) ChatGPT referral incidence by context.
\end{minipage}
\exhibitnotes[-6pt]{Surrounding context is the dominant content category of a household's other browsing in the fifteen minutes before and after a ChatGPT session or Google search session. Sessions with no classifiable surrounding browsing are labeled solo. Panel (a) shows ChatGPT minus Google context shares in percentage points, excluding mixed and other contexts. Panel (b) shows the percentage of ChatGPT sessions with a referral within each displayed context. Error bars show 95\% confidence intervals. \appref{oa:context} provides context-construction details.}
\end{figure}

\autoref{fig:context}, panel~(a), plots the difference in context shares. Reference/knowledge and tools/SaaS contexts account for larger shares of ChatGPT use than of Google use, by 9.0 and 4.3 percentage points, respectively. News/journalism, games and guides, and adult contexts account for smaller shares. ChatGPT sessions are 3.6 percentage points less likely than Google search sessions to be solo. ChatGPT use is thus more often embedded in other browsing than Google search.

Panel~(b) reports ChatGPT's referral incidence within each context. The contexts where use is relatively common differ from those where referrals are most frequent. Referral incidence is highest in developer/technical contexts (13.6\%), academic research (12.9\%), and e-commerce brands (9.5\%), and lowest in adult, social, and reference/knowledge contexts. Reference/knowledge illustrates the difference: it accounts for a relatively larger share of ChatGPT use, yet its referral incidence is among the lowest.\footnote{See \appref{oa:context} for sensitivity to the context-construction rules.}

This section has examined the first channel through which AI search can affect web traffic: when households use ChatGPT, do they click on the external website links it provides, and where do those clicks go? A referral accompanies 5.4\% of ChatGPT sessions, compared with 31.1\% of Google queries, and the gap persists within households and within surrounding contexts. The referrals ChatGPT does send reach reference, academic, developer, and tools/SaaS websites in larger shares than Google's, and social media and ad-supported websites in smaller shares. Pooled across households, ChatGPT referrals are less concentrated than Google's. The household comparisons show no corresponding reduction in concentration, and ChatGPT referrals are more concentrated among household-weeks with more referrals. ChatGPT use is more often embedded in reference and tools browsing than Google search, and its referral incidence varies widely across contexts.

\section{ChatGPT Use and Conventional Search}\label{sec:displacement}

While \autoref{sec:results} examined the first channel through which AI search can affect web traffic, this section examines the second: does AI search change how households use conventional search, such as Google? Answering this requires knowing how households' conventional searches would have evolved without ChatGPT, and no single comparison supplies that counterfactual cleanly. We therefore use two designs. The first, adapted from \citet{padilla2025}, follows households from the week they begin sustained ChatGPT use and compares their conventional searches with those of households that adopt later (\autoref{sec:adoption}). It shows the size and timing of the response to adoption, but households choose when to adopt, for example when their information needs change, so it describes changes that accompany adoption rather than effects of adoption. The second design uses the February~5, 2025 expansion of Search access to households using ChatGPT without an account as a shock whose timing OpenAI set, and estimates how their conventional searches change (\autoref{sec:access-results}).

\subsection{Evidence from household adoption}\label{sec:adoption}

We follow households as they start using ChatGPT and, following \citet{padilla2025}, define adoption as the beginning of sustained use: the first of at least three consecutive complete weeks with a recorded ChatGPT session. This rule distinguishes sustained use from isolated experimentation, while allowing occasional use before adoption. Within the balanced panel, requiring a pre-adoption history of at least four weeks and positive ACS weights yields 1{,}008 households (\appref{oa:lbs-replication}).

\subsubsection{Search declines following ChatGPT adoption}\label{sec:adoption-decline}

Households adopt in different weeks. We therefore use the \citet{callaway2021} estimator to compare each adoption cohort with households that will adopt but have not yet done so. Restricting comparisons to eventual adopters avoids relying on households that never adopt, whose search behavior may differ systematically. Let $G_i$ denote household $i$'s adoption week and $Y_{it}$ its number of Google, Bing, and Yahoo searches in calendar week $t$. For cohort $g$ and a post-adoption week $t\geq g$, we estimate
\begin{equation}
\widehat{ATT}(g,t)
=\overline{Y_{it}-Y_{i,g-1}}^{\,\mathrm{ACS}}_{G_i=g}
-\overline{Y_{it}-Y_{i,g-1}}^{\,\mathrm{ACS}}_{G_i>t},
\label{eq:adoption-att}
\end{equation}
where the bars denote ACS-weighted means. The first term measures the adopting cohort's change from the week before adoption; the second measures the change over the same calendar weeks among households that have yet to adopt. We aggregate cohort estimates at each relative week $w=t-g$, using the contributing cohorts' ACS weight shares. This allows the response to differ across adoption cohorts and over time since adoption.

\begin{figure}[!htbp]
\centering
\caption{Conventional searches rise briefly at ChatGPT adoption and then decline.}
\label{fig:adoption-event}
\includegraphics[width=0.52\linewidth]{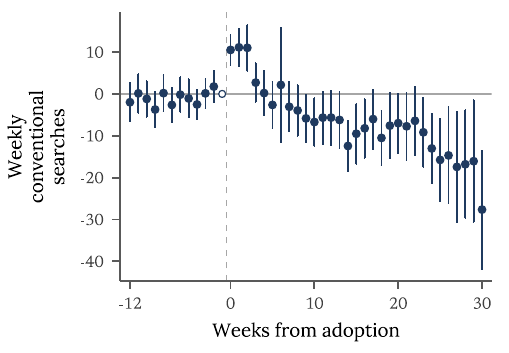}
\exhibitnotes[-6pt]{Points show ACS-weighted \citet{callaway2021} estimates of changes in weekly Google, Bing, and Yahoo searches per household, using not-yet-adopting households as comparisons. The sample includes 1{,}008 eligible adopters. Estimates are relative to the week before ChatGPT adoption, normalized to zero. Error bars show pointwise 95\% normal confidence intervals based on household-clustered standard errors.}
\end{figure}

\autoref{fig:adoption-event} plots the resulting weekly estimates. The pre-adoption estimates show no statistically detectable divergence between the groups (joint test $p=0.295$). Conventional search does not fall immediately when households begin sustained ChatGPT use. Instead, searches rise relative to the comparison households in the adoption week and the following three weeks, then subsequently decline. The subsequent decline is consistent with \citet{padilla2025}, who document a gradual reduction in conventional search following LLM adoption in 2022--2023.

The initial increase could reflect conversations prompting further searches. It could also reflect changes in information needs that lead households to use both services more---for example, when beginning a job search or a new project. Such changes could themselves affect subsequent search activity. The estimates therefore describe how conventional search evolves around sustained ChatGPT adoption; they do not establish that adopting ChatGPT causes the subsequent decline in conventional search.

\subsubsection{ChatGPT Search availability and search substitution}\label{sec:adoption-search-availability}

The rollout of ChatGPT Search provides a comparison that may help distinguish the role of Search availability from that of general LLM use. Search became available to paid subscribers on October~31, 2024, free logged-in users on December~16, and users without an account on February~5, 2025 (\autoref{fig:timeline}). Households beginning sustained use on or after the final expansion could therefore access Search regardless of subscription or login status. Earlier adopters began under more restricted availability, although some already had access and their follow-up also extends beyond February~5. If Search contributes to substitution, conventional searches may decline sooner after adoption when Search is universally available.

\begin{figure}[!htb]
\centering
\captionsetup{skip=2pt}
\caption{ChatGPT use rises after February, while conventional searches decline after March.}
\label{fig:chatgpt-volume}
\begin{minipage}[t]{0.49\linewidth}
\centering
\includegraphics[width=\linewidth]{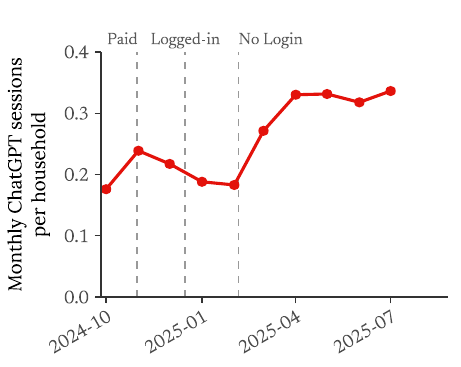}
\smallskip\small (a) ChatGPT sessions.
\end{minipage}\hfill
\begin{minipage}[t]{0.49\linewidth}
\centering
\includegraphics[width=\linewidth]{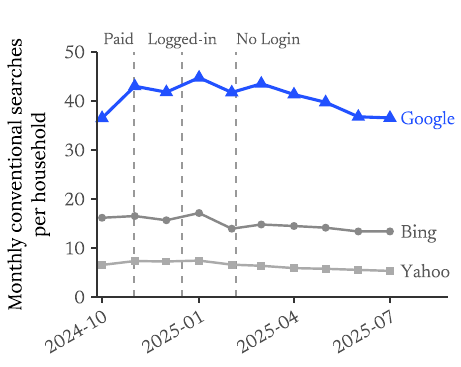}
\smallskip\small (b) Conventional searches.
\end{minipage}
\exhibitnotes[-6pt]{Monthly ChatGPT sessions (panel a) and Google, Bing, and Yahoo searches (panel b) per household in the balanced panel of \panelN{} Comscore U.S. desktop households, October~2024--July~2025. Each monthly total is divided by all \panelN{} households. Dashed lines mark the October~31, December~16, and February~5 ChatGPT Search access expansions.}
\end{figure}

We classify households using the first recorded ChatGPT session day in their adoption week: 366 adopted before February~5 and 642 adopted on or after that date. For each group, we aggregate the cohort-specific estimates described above, retaining not-yet-adopting households as comparisons. Comparing the groups at the same number of weeks since adoption holds the stage of adoption fixed, while Search availability at onset differs. \autoref{fig:chatgpt-volume} provides calendar-time context: monthly ChatGPT sessions per household rise after February, while conventional searches decline from March through July. These aggregate movements do not identify the contribution of Search access.

\autoref{tab:adoption-windows} summarizes the adoption paths over three windows. Weeks 0--3 capture the immediate response, including the adoption week. Weeks 4--19 cover the subsequent period reported for both groups, allowing comparison at the same stage of adoption. Weeks 20--30 describe longer-run changes, for which separate later-adopter estimates are not reported. Each window equally weights its available weekly estimates. The first three columns report the \citet{callaway2021} estimates for all, earlier, and later adopters relative to week $-1$. The final column compares the two groups using generalized method of moments (GMM; \citealp{chen2025efficient}), combining changes measured from four pre-adoption baselines, weeks $-4$ through $-1$.
\begin{table}[!htbp]
\centering
\caption{Conventional searches rise at adoption, then decline sooner among later adopters.}
\label{tab:adoption-windows}
\small
\setlength{\tabcolsep}{4pt}
\begin{tabular}{lrrrr}
\toprule
 & \multicolumn{4}{c}{Outcome: weekly conventional searches per household} \\
\cmidrule(lr){2-2} \cmidrule(lr){3-4} \cmidrule(lr){5-5}
 & All adopters & Earlier adopters & Later adopters & Later minus earlier \\
\midrule
Weeks 0--3 & $8.84^{***}$ & $12.98^{***}$ & $6.17^{**}$ & $-3.68$ \\
 & (1.94) & (3.08) & (2.42) & (2.31) \\
Weeks 4--19 & $-5.73^{**}$ & $-0.52$ & $-14.78^{***}$ & $-13.01^{***}$ \\
 & (2.41) & (2.87) & (4.55) & (3.91) \\
Weeks 20--30 & $-13.80^{***}$ & $-13.72^{***}$ & -- & -- \\
 & (4.64) & (4.64) &  &  \\
\midrule
Observations & 40,320 & 40,320 & 25,680 & -- \\
\addlinespace[5pt]
{\renewcommand{\arraystretch}{0.75}\begin{tabular}[c]{@{}l@{}}Pre-adoption weekly\\conventional searches\end{tabular}} & 35.34 & 37.20 & 34.31 &  \\
\addlinespace[5pt]
Pre-trend test $p$-value & 0.295 & 0.280 & 0.291 & 0.023 \\
\bottomrule
\end{tabular}
\par\smallskip\begin{minipage}{\linewidth}\footnotesize \textit{Notes:} Estimates measure changes in weekly Google, Bing, and Yahoo searches per household using ACS weights. The first three columns report \citet{callaway2021} estimates relative to the week before ChatGPT adoption, with not-yet-adopting households as comparisons; the final column reports the four-baseline GMM later-minus-earlier contrast combining baselines from weeks $-4$ through $-1$. Earlier and later adopters are classified by whether their first ChatGPT session in the adoption week occurs before or on/after February~5, 2025. The GMM pre-trend test includes week $-1$, which is estimated rather than normalized to zero. \appref{oa:adoption-gmm} reports GMM diagnostics. Standard errors in parentheses are household-clustered in the first three columns and based on 999 household bootstrap refits in the final column. $^{*}p<0.10$, $^{**}p<0.05$, $^{***}p<0.01$.\end{minipage}
\end{table}

Across all adopters, the initial increase gives way to a substantial decline in conventional search. By weeks 20--30, the estimated reduction is equivalent to about 39\% of adopters' pre-adoption weekly search activity. The group comparison shows that this decline emerges sooner among later adopters. During weeks 4--19, their conventional searches are already substantially below the path implied by the comparison households, whereas the earlier-adopter estimate remains close to zero. Earlier adopters reach a similar decline in weeks 20--30. The difference is therefore principally in when the decline appears.

\begin{figure}[!htbp]
\centering
\caption{Conventional searches decline sooner among later adopters.}
\label{fig:adoption-timing}
\begin{minipage}[c]{0.04\linewidth}\centering\rotatebox{90}{\small Weekly conventional searches}\end{minipage}%
\begin{minipage}[c]{0.46\linewidth}\includegraphics[width=\linewidth]{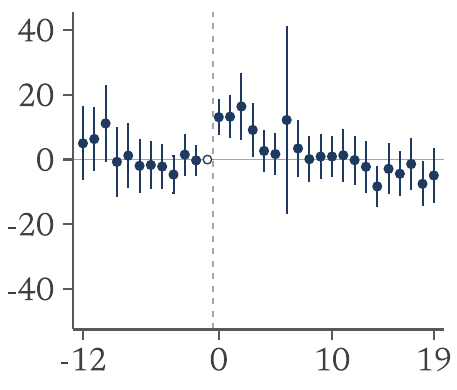}\end{minipage}%
\hfill
\begin{minipage}[c]{0.46\linewidth}\includegraphics[width=\linewidth]{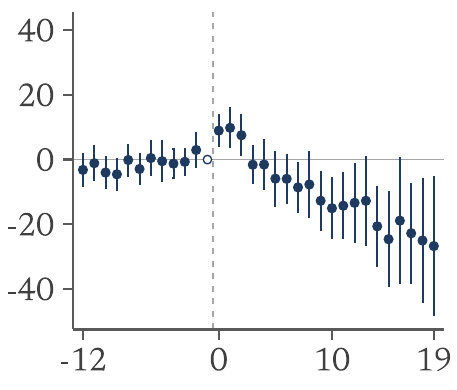}\end{minipage}
\par\smallskip
\begin{minipage}[t]{0.04\linewidth}\mbox{}\end{minipage}%
\begin{minipage}[t]{0.46\linewidth}\centering Weeks from adoption\par\smallskip\small (a) Earlier adopters.\end{minipage}%
\hfill
\begin{minipage}[t]{0.46\linewidth}\centering Weeks from adoption\par\smallskip\small (b) Later adopters.\end{minipage}
\par\bigskip
\begin{minipage}[c]{0.04\linewidth}\centering\rotatebox{90}{\small Weekly conventional searches}\end{minipage}%
\begin{minipage}[c]{0.46\linewidth}\includegraphics[width=\linewidth]{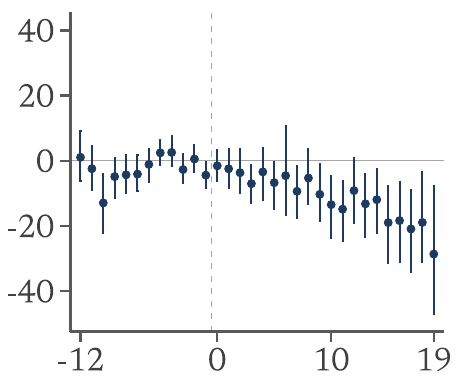}\end{minipage}%
\hspace*{0.04\linewidth}
\par\smallskip
\begin{minipage}[t]{0.04\linewidth}\mbox{}\end{minipage}%
\begin{minipage}[t]{0.46\linewidth}\centering Weeks from adoption\par\smallskip\small (c) Later minus earlier.\end{minipage}%
\hspace*{0.04\linewidth}
\exhibitnotes{Panels (a) and (b) show ACS-weighted \citet{callaway2021} estimates of changes in weekly Google, Bing, and Yahoo searches per household, using not-yet-adopting households as comparisons. Earlier and later adopters are classified by whether their first ChatGPT session in the adoption week occurs before or on/after February~5, 2025. Panel (c) reports the later-minus-earlier contrast using four-baseline GMM, combining weeks $-4$ through $-1$ (\citealp{chen2025efficient}). Week $-1$ is normalized to zero in panels (a)--(b) and estimated using the other three baselines in panel (c). Error bars show pointwise 95\% confidence intervals: household-clustered normal intervals in panels (a)--(b), and basic intervals from 999 household bootstrap refits in panel (c).}
\end{figure}

\autoref{fig:adoption-timing} makes this timing difference visible. Panels~(a) and~(b) trace the earlier- and later-adopter responses at the same number of weeks since adoption. Panel~(c) compares those responses directly using GMM: values below zero indicate a more negative response among later adopters. Over weeks 4--19, the estimated later-adopter response is about 13 searches per household-week more negative than the earlier-adopter response.\footnote{The GMM contrast's joint pre-adoption test rejects zero ($p=0.023$), and its inference is exploratory because the estimator was selected after comparing alternatives. \appref{oa:adoption-gmm} reports these diagnostics and \appref{fig:oa-adoption-original-contrast} the corresponding single-baseline comparison.} The evidence thus points to an earlier decline under broader Search availability; the shorter follow-up for later adopters prevents a comparison of the groups' eventual declines.

The earlier decline among later adopters is consistent with broader Search availability contributing to conventional-search substitution. However, households choose when to adopt ChatGPT and may do so when their information needs increase. Those changing needs can themselves affect conventional search, making it difficult to attribute the observed changes to ChatGPT. \autoref{sec:access-results} therefore turns to the February expansion of Search access, whose timing was set by OpenAI.

\subsection{Evidence from the Search access expansion}\label{sec:access-results}

On February~5, 2025, OpenAI expanded ChatGPT Search to users without an account \citep{openai2024search}. Households already using ChatGPT anonymously gained access on a common date that they could not choose. The expansion thus provides a change in Search availability whose timing is external to individual households' adoption decisions. We use this change to examine whether expanded Search access increases ChatGPT use and leads households to substitute it for conventional search.

\subsubsection{Overall effects on ChatGPT use and conventional search}\label{sec:access-average}

The February expansion made Search available to all ChatGPT users, leaving no within-ChatGPT group without Search access. We therefore compare 945 households that had used ChatGPT anonymously before the expansion with a comparison group of 429 households that had used Gemini but had not used ChatGPT. Gemini users provide a useful comparison group because they, too, had already adopted an AI assistant. Their behavior therefore helps account for changes in conventional search among households already engaging with AI during the same period. The comparison asks whether prior-anonymous ChatGPT users change their behavior differently after gaining Search access.

We estimate the change in ChatGPT sessions and in conventional searches, using the 18 weeks before and 25 weeks after the expansion. For each outcome $Y$, the estimate in relative week $w$ is
\begin{equation}
\widehat{\delta}_{w}
=\left(\overline{Y}_{T,w}-\overline{Y}_{C,w}\right)
-\left(\overline{Y}_{T,-1}-\overline{Y}_{C,-1}\right),
\label{eq:access-att}
\end{equation}
where $T$ denotes prior-anonymous ChatGPT households, $C$ the Gemini comparison group, and bars ACS-weighted means. The estimate measures how much the gap between the groups changes relative to the week before the expansion. On our balanced panel, these estimates equal those from an event-study regression with household and calendar-week fixed effects. We average the weekly estimates over weeks 0--24 to summarize the post-expansion response.\footnote{See \appref{oa:cohort-schema} for sample construction and \appref{oa:master-table} for the conventional-search outcome cap and household-level inference.}

\autoref{tab:february-access} reports these averages and the pre-expansion diagnostics. The joint tests do not detect pre-expansion divergence in either ChatGPT sessions or conventional searches. After the expansion, ChatGPT use increases, but this additional use is not accompanied by a detectable overall decline in conventional search. The conventional-search estimate is close to zero, although its uncertainty does not rule out increases or decreases.
\begin{table}[!htbp]
\centering
\caption{ChatGPT use rises without detected declines in conventional search.}
\label{tab:february-access}
\small
\setlength{\tabcolsep}{4pt}
\begin{tabular}{lcc}
\toprule
 & \multicolumn{2}{c}{Outcomes: weekly counts per household} \\
\cmidrule(lr){2-3}
 & ChatGPT sessions & Conventional searches \\
\midrule
Average post-expansion change & $0.117^{***}$ & $0.00$ \\
 & (0.022) & (1.88) \\
\midrule
Observations & 59,082 & 59,082 \\
Pre-trend test $p$-value & 0.433 & 0.248 \\
\bottomrule
\end{tabular}
\par\smallskip\begin{minipage}{\linewidth}\footnotesize \textit{Notes:} Estimates compare 945 households that previously used ChatGPT anonymously with 429 Gemini-using households with no prior ChatGPT use. Estimates measure the average change after the February~5, 2025 expansion relative to the preceding week, using household and calendar-week fixed effects and ACS weights. Conventional searches count Google, Bing, and Yahoo result-page loads and are winsorized at the 99th percentile. Standard errors from 999 household bootstrap draws appear in parentheses. \appref{oa:master-table} reports identification and inference details; \appref{tab:oa-access-gmm} reports corresponding GMM estimates. $^{*}p<0.10$, $^{**}p<0.05$, $^{***}p<0.01$.\end{minipage}
\end{table}

\begin{figure}[!htbp]
\centering
\caption{ChatGPT use rises without detected declines in conventional search.}
\label{fig:february-access}
\begin{minipage}[t]{0.49\linewidth}
\centering
\includegraphics[width=\linewidth]{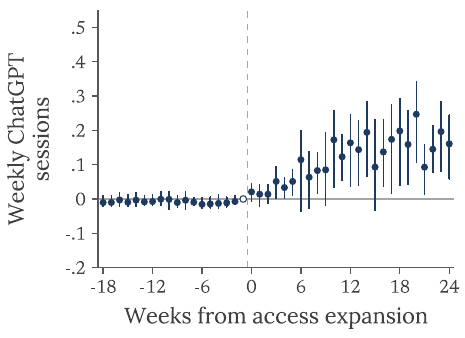}
\smallskip\small (a) ChatGPT sessions.
\end{minipage}\hfill
\begin{minipage}[t]{0.49\linewidth}
\centering
\includegraphics[width=\linewidth]{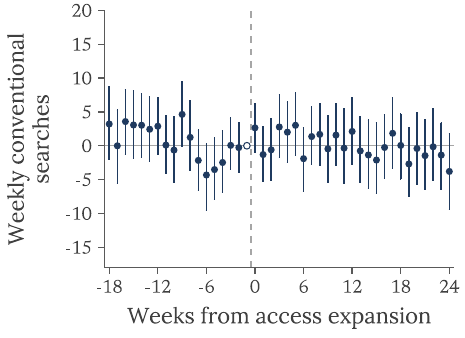}
\smallskip\small (b) Conventional searches.
\end{minipage}
\exhibitnotes[-6pt]{Points show ACS-weighted TWFE estimates comparing 945 prior-anonymous ChatGPT households with 429 prior-Gemini households, with household and calendar-week fixed effects (\appref{oa:master-table}). Outcomes are ChatGPT sessions and Google, Bing, and Yahoo searches per household-week. Open circles mark week $-1$, normalized to zero. Bars show pointwise 95\% basic intervals from 999 household bootstrap draws. Dashed lines mark the February~5 expansion.}
\end{figure}

The averages summarize the response over the post-expansion period. \autoref{fig:february-access} shows how that response evolves week by week, plotting the estimates from \autoref{eq:access-att}. The ChatGPT session estimates rise over the weeks following the expansion, while the conventional-search estimates fluctuate around zero with intervals that include zero. The weekly paths thus show growing ChatGPT use without a sustained, statistically detectable overall decline in conventional search.

\subsubsection{Heterogeneity by prior search activity}\label{sec:prior-search}

We find no detectable average decline in conventional search following the February expansion, but this average may conceal different responses across households. Prior search activity captures how heavily households rely on conventional search and may also reflect differences in the needs it serves. Households that search frequently have more searches that ChatGPT could potentially replace, but their searches may also serve needs that are harder to address within a conversation. We therefore examine whether the response to expanded Search access differs with pre-expansion search activity.

We divide households by their mean weekly conventional searches over the 18 pre-expansion weeks, splitting at the treated households' weighted median of 18.5 (\appref{oa:cohort-schema}). Within each group, we repeat the comparison in \autoref{eq:access-att}, using prior Gemini users in the same activity group as comparisons. \autoref{tab:prior-search} reports the average responses over weeks 0--24; \appref{oa:prior-search-event} reports the weekly conventional-search estimates.

\begin{table}[!htbp]
\centering
\caption{Conventional search declines among households with lower prior activity.}
\label{tab:prior-search}
\small
\setlength{\tabcolsep}{4pt}
\begin{tabular}{lcccc}
\toprule
 & \multicolumn{4}{c}{Outcomes: weekly counts per household} \\
\cmidrule(lr){2-5}
 & \multicolumn{2}{c}{Lower prior search activity} & \multicolumn{2}{c}{Higher prior search activity} \\
\cmidrule(lr){2-3} \cmidrule(lr){4-5}
 & \begin{tabular}[c]{@{}c@{}}ChatGPT\\sessions\end{tabular} & \begin{tabular}[c]{@{}c@{}}Conventional\\searches\end{tabular} & \begin{tabular}[c]{@{}c@{}}ChatGPT\\sessions\end{tabular} & \begin{tabular}[c]{@{}c@{}}Conventional\\searches\end{tabular} \\
\midrule
Average post-expansion change & $0.109^{***}$ & $-3.71^{**}$ & $0.127^{***}$ & $2.88$ \\
 & (0.024) & (1.35) & (0.035) & (3.20) \\
\midrule
Observations & 29,842 & 29,842 & 29,240 & 29,240 \\
Pre-trend test $p$-value & 0.408 & 0.293 & 0.417 & 0.405 \\
\bottomrule
\end{tabular}
\par\smallskip\begin{minipage}{\linewidth}\footnotesize \textit{Notes:} Households are divided by their mean weekly conventional searches before the expansion, using the treated households' ACS-weighted median as the cutoff. Within each group, estimates compare households that previously used ChatGPT anonymously with Gemini-using households with no prior ChatGPT use. Estimates measure the average change after the February~5, 2025 expansion relative to the preceding week, using the same specification and outcome definitions as \autoref{tab:february-access}. Standard errors from 999 household bootstrap draws appear in parentheses. \appref{oa:cohort-schema} reports group construction; \appref{tab:oa-access-gmm} reports corresponding GMM estimates. $^{*}p<0.10$, $^{**}p<0.05$, $^{***}p<0.01$.\end{minipage}
\end{table}

ChatGPT use rises by similar amounts in both groups, but conventional search declines among households with lower prior search activity. Their estimated reduction is about four searches per household per week, equivalent to roughly 36\% of their pre-expansion search activity. For higher-activity households, the estimate is positive and imprecise, so we do not detect either an increase or a decline. The direct comparison confirms that the groups respond differently: the lower-minus-higher difference is $-6.59$ searches per household-week ($p=0.038$). Similar average increases in ChatGPT sessions therefore accompany different conventional-search responses. The evidence suggests that expanded Search access led households with lower prior search activity to substitute ChatGPT for some conventional search.

\subsubsection{The role of search use cases}\label{sec:traffic}

Could differences in the needs served by conventional search help explain these household responses? As \autoref{sec:framework} suggests, an answer may replace some searches more readily than others. We examine this possibility by first comparing where the two groups' conventional-search referrals led before the expansion, and then estimating how referrals to different destination categories changed after it.

\begin{figure}[!htbp]
\centering
\caption{Conventional-search referral patterns differ by prior activity and destination.}
\label{fig:prior-search-composition}
\label{fig:content-heterogeneity}
\begin{minipage}[b]{0.495\linewidth}
\centering
\includegraphics[width=\linewidth]{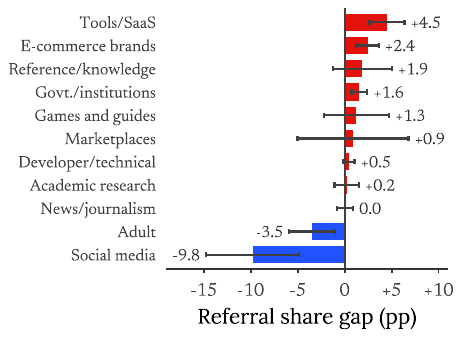}
\smallskip\small (a) Lower minus higher prior-search households: referral share.
\end{minipage}
\hfill
\begin{minipage}[b]{0.495\linewidth}
\centering
\includegraphics[width=\linewidth]{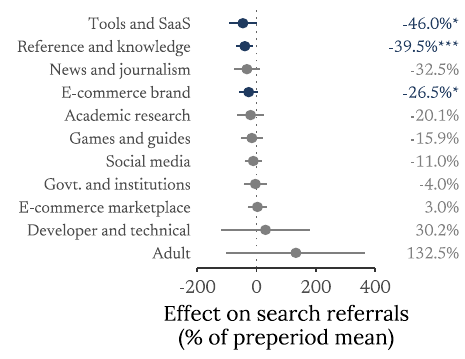}
\smallskip\small (b) Referral effects by category.
\end{minipage}
\exhibitnotes[-6pt]{Both panels use ACS-weighted referrals from Google, Bing, and Yahoo to eleven destination categories. Panel (a) shows pre-expansion referral shares for lower- minus higher-prior-search treated households, in percentage points. Panel (b) reports average post-expansion TWFE effects for the full sample, comparing prior-anonymous ChatGPT households with Gemini comparison households, with household and calendar-week fixed effects and the week before expansion as the reference. Effects are expressed as percentages of treated households' pre-expansion mean referrals in each category. Intervals are pointwise 95\% normal confidence intervals based on household-clustered standard errors. Dark points in panel (b) indicate $p<0.10$; $^{*}p<0.10$, $^{**}p<0.05$, $^{***}p<0.01$. \appref{oa:category-outcomes} provides construction details and GMM comparisons.}
\end{figure}

For the first comparison, we use referrals from Google, Bing, and Yahoo among the treated households over the 18 pre-expansion weeks. We classify their destinations into eleven content categories and calculate each category's share of referrals separately for lower- and higher-activity households. \autoref{fig:prior-search-composition}, panel~(a), plots the lower-minus-higher share difference: a positive value means the category accounts for a larger share of lower-activity households' referrals.

Lower-activity households direct a larger share of their referrals to tools/SaaS websites and a smaller share to social media, by 4.5 and 9.8 percentage points, respectively (Holm-adjusted $p<0.001$ for both). They also direct larger shares to e-commerce brand websites and smaller shares to adult websites. These differences suggest that the groups use conventional search for different purposes. Tools such as Microsoft Math Solver serve needs that a ChatGPT conversation might address directly. Search-enabled answers may similarly address product questions that otherwise lead users to brand websites, such as Apple's iPhone pages. By contrast, watching a YouTube video or participating in a Reddit discussion requires reaching the platform itself.

For the second comparison, we apply the access-expansion design to weekly conventional-search referral counts in each destination category. These estimates use the full access sample, comparing all 945 prior-anonymous ChatGPT households with the 429 Gemini comparison households. We average the weekly estimates over weeks 0--24 and express each as a percentage of the treated households' pre-expansion mean for that category (\appref{oa:category-outcomes}). \autoref{fig:prior-search-composition}(b) presents these category estimates for the full access sample.

\autoref{fig:prior-search-composition}(b) shows estimated declines in conventional-search referrals to tools/SaaS and e-commerce brand websites of about 46\% and 27\%, respectively, with both estimates significant at the 10\% level. Neither social-media nor adult referrals show a statistically detectable change. The declines thus occur in two categories that account for larger shares of lower-activity households' referrals before the expansion.\footnote{We do not interpret the estimated decline for reference/knowledge because treated and comparison households already diverged before the expansion (joint lead-test $p=0.009$; \appref{oa:category-outcomes}).} The category estimates are sensitive to specification and come from the full access sample, so they suggest a possible explanation for the household differences rather than establish which search needs drive them.\footnote{The appendix compares TWFE with GMM for the ten categories excluding adult websites. Under GMM, the tools/SaaS and brand estimates are smaller and all ten full-sample category intervals include zero (\appref{oa:category-outcomes}).}

Expanded Search access increases ChatGPT use without a detectable overall decline in conventional search. That average conceals different responses: households with lower prior search activity reduce conventional search, while no decline is detected among higher-activity households, despite similar increases in ChatGPT sessions. The referral evidence suggests that differences in search needs may help explain this pattern, with substitution more plausible for some task-oriented and product-related searches.

\section{Discussion}\label{sec:conclusion}

AI search changes how users reach websites through two channels, and the evidence here separates them. For the first channel, observing use alongside referrals shows that ChatGPT sends users onward much less often than Google: a referral accompanies 5.4\% of ChatGPT sessions, compared with 31.1\% of Google queries, and its referrals give ad-supported websites a smaller share and reference, technical, and tool websites larger shares. For the second channel, conventional searches (Google, Bing, and Yahoo) decline after households adopt ChatGPT. The decline comes sooner among households that adopt from February~5, 2025, when ChatGPT Search became available without an account. Among households already using ChatGPT without an account, conventional searches show no detectable overall change after that expansion, but fall for those with lower prior search activity.

Both channels are consistent with a single boundary: whether an answer can meet the need behind a visit or a search, or only the website can. Tools/SaaS websites illustrate it from both sides. Conventional-search referrals to these websites fall after the access expansion (significant at the 10\% level), as answers may take over tasks such as a calculation; yet tools account for a larger share of the referrals ChatGPT itself makes, plausibly when the tool must be used on its website. Searches that lead to platform content or interaction, such as social media, show no detectable decline. The boundary can therefore run through categories of websites rather than between them; where it falls for a given website depends on what its users need from it.

For research on information intermediaries, these findings suggest that AI search loosens the connection between supplying information and receiving visits on which websites have relied \citep{dellarocas2013}. Content aggregators raised a similar tension between substitution and discovery in news \citep{dellarocas2016,chiou2017,calzada2020}. AI search extends it to answers across many tasks and adds a second margin: beyond referring users less often, it can reduce the conventional searches that would otherwise have led them to websites. Its referrals are also less concentrated than Google's in aggregate. The household comparisons show no corresponding reduction in concentration, and ChatGPT referrals are more concentrated among household-weeks with more referrals. This contrast is consistent with different households directing their referrals to different websites.

For websites, the two channels imply that overall traffic depends on AI referrals and conventional-search referrals together. A website can rank highly among ChatGPT's destinations yet lose visits overall if declines in conventional-search referrals outweigh its gains from AI referrals. Because conventional searches lead to external visits far more often than ChatGPT conversations do, such losses are most likely where an answer can deliver what users came for, and least likely where a website's value lies in interaction, participation, or transactions. Assessing the economic consequences therefore requires measuring both channels and how arriving visits contribute to advertising revenue, subscriptions, transactions, or participation. These measures also bear on policy. The European Commission designates ChatGPT a search engine (\autoref{sec:intro}), consistent with the intermediary role documented here; our evidence adds that, as used, this search engine sends users to websites far less often than Google does.

The two channels together raise a longer-run question about the returns to producing the information that AI search uses. They motivate examining the relationship between AI intermediaries and information providers alongside research on publisher responses to generative AI \citep{zhao2026} and online knowledge sharing \citep{burtch2024,delriochanona2024}. Linking retrieval and citation records to referrals, publisher revenues, and content investment could clarify how supplying information to AI systems translates into returns to its producers. The central question for the next generation of information intermediaries is whether they can improve access to web information while sustaining the incentives to produce and maintain it.

\bibliography{references}

\clearpage
\setcounter{section}{0}
\setcounter{figure}{0}
\setcounter{table}{0}
\setcounter{equation}{0}
\renewcommand{\theHsection}{OA.\arabic{section}}
\renewcommand{\thesection}{OA.\arabic{section}}
\renewcommand{\thesubsection}{\thesection.\arabic{subsection}}
\renewcommand{\thesubsubsection}{\thesubsection.\arabic{subsubsection}}
\renewcommand{\thefigure}{\thesection.\arabic{figure}}
\renewcommand{\thetable}{\thesection.\arabic{table}}
\renewcommand{\theequation}{\thesection.\arabic{equation}}
\makeatletter
\renewcommand{\l@section}{\@dottedtocline{1}{0em}{4.2em}}
\renewcommand{\l@subsection}{\@dottedtocline{2}{1.5em}{5.0em}}
\renewcommand{\l@subsubsection}{\@dottedtocline{3}{3.8em}{5.8em}}
\makeatother
\makeatletter
\@addtoreset{figure}{section}
\@addtoreset{table}{section}
\@addtoreset{equation}{section}
\makeatother
\setlength{\emergencystretch}{3em}
\singlespacing

\setcounter{topnumber}{2}
\setcounter{bottomnumber}{1}
\setcounter{totalnumber}{3}
\renewcommand{\topfraction}{0.88}
\renewcommand{\bottomfraction}{0.75}
\renewcommand{\textfraction}{0.08}
\renewcommand{\floatpagefraction}{0.75}
\setlength{\textfloatsep}{14pt plus 2pt minus 3pt}
\setlength{\floatsep}{14pt plus 2pt minus 3pt}

\manuscripttitle{Online Appendix}{AI Search as an Information Intermediary:\\Clickstream Evidence on Web Traffic}
\thispagestyle{empty}

\noindent This appendix documents how the measures in the main paper are constructed and validated, and reports supplementary estimates. The guide below links each part to the main paper.

\begin{center}
\small
\begin{tabular}{@{}p{0.45\linewidth}p{0.48\linewidth}@{}}
\toprule
Main-paper question & Supporting material \\
\midrule
\multicolumn{2}{@{}l}{\emph{Data used throughout}} \\
What is observed and reconstructed? & \autoref{oa:data}: visits, panel, request paths, sessions \\
\addlinespace
\multicolumn{2}{@{}l}{\emph{Referrals from ChatGPT (main-paper \autoref{sec:results})}} \\
How often does use produce a referral? & \autoref{oa:ctr}: definition, attribution, and checks; \autoref{oa:context}: surrounding browsing \\
Which websites receive referrals? & \autoref{oa:classification}: categories and validation; \autoref{oa:concentration}: concentration and prominence \\
\addlinespace
\multicolumn{2}{@{}l}{\emph{Conventional search (main-paper \autoref{sec:displacement})}} \\
How do conventional searches change after adoption? & \autoref{oa:causal}: measures, sample, estimator, earlier and later adopters \\
How do they change after the February access expansion? & \autoref{oa:access}: households, estimation, GMM comparison, referral categories \\
\bottomrule
\end{tabular}
\end{center}
\clearpage
\clearpage
\section{Data construction and observable activity}\label{oa:data}

The Comscore data record the web requests made from each household's computer. This section explains which requests we count as webpage visits, which households form the balanced panel, and how ChatGPT's request paths reveal conversations and messages. Descriptive analyses use all active households; analyses of conventional search use the balanced panel.

\subsection{Foreground visits and background requests}\label{oa:cleaning}

A browser issues many requests for each page a person opens: the page itself, and then scripts, images, and data. We count a record as a foreground visit when its content type is \texttt{text/\allowbreak{}html} and its HTTP response code indicates a successful or redirected request.\footnote{The accepted codes are 200, 201, 202, 203, 204, 206, 301, 302, 303, 304, 307, and 308.} This keeps page loads and drops scripts, images, fonts, audio and video, JSON responses, event streams, document downloads, and records with error codes or malformed content types.

Some of ChatGPT's background requests carry information that page loads do not: they record when a conversation is opened, when a message is sent, and whether the user is logged in. We use these requests to measure ChatGPT activity (\autoref{oa:endpoint-coverage}), and we count only foreground visits as webpage visits. Referrals apply further destination exclusions (\autoref{oa:referral-def}).

\subsection{Balanced panel and activity threshold}\label{oa:panel-threshold}

The balanced panel contains \panelN{} households with at least four foreground visits in each month from October~2024 through July~2025. Comscore's machine identifier defines a household. The threshold of four asks for repeated evidence that a household is observed in every month, and it keeps lightly active households in the panel. A threshold of one would add 4{,}822 households, and a threshold of fifty would remove 15{,}009 (\autoref{tab:oa-panel-threshold}).

\begin{table}[!htbp]
  \centering
  \caption{The monthly activity threshold changes panel coverage.}
  \label{tab:oa-panel-threshold}
  \footnotesize

  \begin{tabular}{lrrrrr}
    \toprule
    Minimum monthly visits & Households & Added & Removed & Mean visits per month & CV \\
    \midrule
    1   & 50{,}208 & $+$4{,}822  & 0       & 577.1 & 1.084 \\
    2   & 47{,}967 & $+$2{,}581  & 0       & 562.8 & 1.061 \\
    3   & 46{,}477 & $+$1{,}091  & 0       & 606.3 & 1.042 \\
    4 (main) & 45{,}386 & 0 & 0 & --- & --- \\
    5   & 44{,}403 & 0 & $-$983      & 512.3 & 1.036 \\
    10  & 41{,}105 & 0 & $-$4{,}281  & 571.9 & 1.010 \\
    20  & 37{,}057 & 0 & $-$8{,}329  & 601.4 & 0.973 \\
    50  & 30{,}377 & 0 & $-$15{,}009 & 652.0 & 0.922 \\
    \bottomrule
  \end{tabular}
  \exhibitnotes{The starting population comprises 50{,}208 households with at least one foreground visit in every sample month. Each row requires the stated minimum number of foreground visits per month. Added and removed households are measured relative to the threshold of four used throughout. Mean monthly activity and its cross-month coefficient of variation (CV) describe the households added or removed at each threshold.}
\end{table}

\subsection{Endpoint definitions and coverage}\label{oa:endpoint-coverage}

Comscore records the host and path of each request. It does not record prompts, responses, or authorization headers. We therefore identify ChatGPT activity from the paths that the web application requests (\autoref{tab:oa-endpoint-defs}), and we interpret those paths with the help of public descriptions of the application.\footnote{See the community-maintained \emph{everything-chatgpt} catalogue, \url{https://github.com/terminalcommandnewsletter/everything-chatgpt}, and the archived anonymous client, \url{https://github.com/Mr-Destructive/anonymous-chatgpt} (accessed 2026-06-24).} Opening a conversation requests a path that contains the conversation's identifier (UUID), and sending a message posts to a separate path. Paths that begin with \texttt{backend-api} serve logged-in users, and paths that begin with \texttt{backend-anon} serve anonymous users. A logged-in user's browser can also issue some anonymous requests when a page first loads, so the access analysis classifies households with the more selective rules in \autoref{oa:cohort-schema}.

\begin{table}[H]
  \centering
  \caption{Endpoint paths distinguish several observable forms of ChatGPT activity.}
  \label{tab:oa-endpoint-defs}
  \footnotesize

  \begin{tabular}{@{}>{\raggedright\arraybackslash}p{0.36\linewidth} >{\raggedright\arraybackslash}p{0.58\linewidth}@{}}
    \toprule
    Endpoint & Definition and role in the paper \\
    \midrule
    \texttt{chatgpt.com} foreground page load &
      GET. A user-visible page load (homepage or
      \texttt{/c/\{uuid\}}); the broadest but shallowest signal.
      \emph{Visit denominator.} \\
    \addlinespace
    \texttt{backend-api/\allowbreak{}conversation/\allowbreak{}\{uuid\}} &
      GET. The browser opening or reloading one conversation; each
      distinct UUID is one conversation. \emph{Conversation
      denominator.} \\
    \texttt{backend-anon/\allowbreak{}conversation/\allowbreak{}\{uuid\}} &
      GET. The same for anonymous sessions; negligible volume. \\
    \addlinespace
    \texttt{backend-api/\allowbreak{}f/\allowbreak{}conversation}\tablefootnote{Unlike the
      other endpoints here, the \texttt{f/\allowbreak{}conversation} variant is
      not separately documented in public reverse-engineering
      sources. When we inspected the web application's network
      requests, sending one message issued exactly one
      \texttt{f/\allowbreak{}conversation} POST. As an
      undocumented internal endpoint it may change without notice.} &
      POST. One logged-in message send (a newer variant
      whose volume grows from March to July~2025). \emph{Message-level referral
      attribution.} \\
    \texttt{backend-anon/\allowbreak{}f/\allowbreak{}conversation} &
      POST. One anonymous message send. \\
    \bottomrule
  \end{tabular}
  \exhibitnotes{The table gives our interpretation of each request path. Conversation loads include returns and reloads; they are grouped into sessions using the rule in \autoref{oa:conv-cleaning}. Conversation loads record conversations with and without web retrieval.}
\end{table}

Coverage differs across these measures (\autoref{tab:oa-endpoint-coverage}). In the balanced panel, 9{,}481 households have a foreground ChatGPT visit during the ten months, and 5{,}265 of them load a logged-in conversation. Message sends are well covered only in July~2025. We therefore build the main measure of use from conversation loads and use message sends for a supplementary July check (\autoref{oa:referral-def}).

\begin{table}[!htbp]
  \centering
  \caption{Endpoint coverage differs across activity measures.}
  \label{tab:oa-endpoint-coverage}
  \footnotesize

  \begin{tabular}{lrrr}
    \toprule
    Endpoint & Households & \% of panel & \% of visitors \\
    \midrule
    \texttt{chatgpt.com} foreground visits          & 9{,}481 & 20.9 & 100.0 \\
    \midrule
    \texttt{backend-api/\allowbreak{}conversation/\allowbreak{}\{uuid\}}      & 5{,}265 & 11.6 & 55.5 \\
    \texttt{backend-anon/\allowbreak{}conversation/\allowbreak{}\{uuid\}}     &     348 &  0.8 &  3.7 \\
    \midrule
    \texttt{backend-api/\allowbreak{}f/\allowbreak{}conversation} (Jul 2025)  & 2{,}361 &  5.2 & 65.8 (Jul) \\
    \texttt{backend-anon/\allowbreak{}f/\allowbreak{}conversation} (Jul 2025) &     936 &  2.1 & 26.1 (Jul) \\
    \bottomrule
  \end{tabular}
  \exhibitnotes{The population is the \panelN-household balanced panel. The first three rows count households observed at any point in the ten months; the message rows report July~2025 and include both sends and preparatory requests. The panel percentage divides by the \panelN{} panel households. The visitor percentage divides by the households with a foreground ChatGPT visit in the same period: 9{,}481 over ten months, and the July count for the July rows.}
\end{table}

\subsection{ChatGPT sessions and gap sensitivity}\label{oa:conv-cleaning}

Our unit of ChatGPT use is the ChatGPT session. A session groups a household's loads of one conversation, whether logged in or anonymous, and a new session starts when more than one hour passes between loads of that conversation. Loads of different conversations always belong to different sessions, and sessions are built within calendar months. The rule treats quick reloads as one occasion of use and a return to an old conversation after a break as a new one.

The gaps between loads motivate the one-hour boundary. When a household loads the same conversation twice in a row, the median gap is thirteen seconds. When it loads another conversation in between, the median gap is 4{,}946 seconds, about 82 minutes. The full sample contains 263{,}008 conversations within household-months and 409{,}133 sessions under the one-hour rule. A thirty-minute rule gives 4.9\% more sessions and a two-hour rule 4.4\% fewer (\autoref{tab:oa-gap-sweep}). Main-paper \autoref{fig:absorption-ratio} repeats the incidence calculation with thirty-minute and five-hour boundaries, and ChatGPT's monthly referral incidence changes little.

\begin{table}[!htbp]
  \centering
  \caption{Session counts vary moderately around the one-hour rule.}
  \label{tab:oa-gap-sweep}
  \footnotesize

  \begin{tabular}{@{}lrr@{}}
    \toprule
    Same-UUID gap & Sessions & vs.\ $1$~h \\
    \midrule
    $0.25$~h ($900$~s)              & 449{,}391 & $+9.8\%$ \\
    $0.5$~h ($1{,}800$~s)           & 429{,}074 & $+4.9\%$ \\
    $1$~h ($3{,}600$~s, main)  & 409{,}133 & --- \\
    $2$~h ($7{,}200$~s)             & 391{,}282 & $-4.4\%$ \\
    \midrule
    Conversations within household-months & 263{,}008 & invariant \\
    \bottomrule
  \end{tabular}
  \exhibitnotes{The sample includes all observed conversation loads in the full October~2024--July~2025 descriptive sample. A new session starts when the gap between a household's loads of the same conversation exceeds the stated threshold. The final column reports the percentage difference from the one-hour count. The final row counts distinct conversations within household-months, which does not depend on the threshold.}
\end{table}

\subsection{Growth in reach and activity per household}\label{oa:decomp}

ChatGPT sessions per active household grew 3.0-fold between October~2024 and July~2025. Most of that growth came from more households using ChatGPT, and less of it from more sessions per user. To separate the two, let $M_t$ be the number of active households in month $t$, $N_t$ the number with a session, $a_t=N_t/M_t$ their share (reach), and $\bar q_t$ their mean session count. Sessions per active household are $v_t=a_t\bar q_t$, and between baseline month $b$ and end month $e$,
\[
\Delta v=\underbrace{(a_e-a_b)\bar q_b}_{\text{reach}}
+\underbrace{a_b(\bar q_e-\bar q_b)}_{\text{activity per reached household}}
+\underbrace{(a_e-a_b)(\bar q_e-\bar q_b)}_{\text{interaction}}.
\]
Reach accounts for 65.5\% of the growth, activity per reached household for 14.9\%, and their interaction for 19.6\% (\autoref{tab:oa-decomp}). Dividing by $M_t$ keeps changes in the size of the active sample out of the comparison. The decomposition uses the full descriptive sample; main-paper \autoref{fig:chatgpt-volume} holds the household sample fixed in the balanced panel.

\begin{table}[!htbp]
  \centering
  \caption{Expanding reach accounts for most measured activity growth.}
  \label{tab:oa-decomp}
  \small

  \begin{tabular}{lr}
    \toprule
 & ChatGPT sessions \\
    \midrule
Start / end month & Oct 2024 / Jul 2025 \\
Growth in volume per household & 3.0-fold \\
Reach share & 65.5\% \\
Activity-per-household share & 14.9\% \\
Interaction share & 19.6\% \\
    \bottomrule
  \end{tabular}
  \exhibitnotes{The decomposition uses the full descriptive sample of ChatGPT sessions from October~2024 to July~2025. Each share divides the corresponding term by the total change in session volume per active household.}
\end{table}

\section{Referral construction and incidence comparisons}\label{oa:ctr}

Main-paper \autoref{sec:absorption} compares how often ChatGPT sessions and Google queries are accompanied by a visit to an external website. This section defines a referral, explains how each referral is assigned to a session or a query, and reports checks on the measure. It then describes the within-household comparisons and relates our estimates to published traffic statistics.

\subsection{Referral definition, attribution, and checks}\label{oa:referral-def}

A referral is a foreground visit to an external website whose HTTP referrer names the intermediary the user came from. A clean referral is one that passes the destination exclusions below. Referral incidence is the share of ChatGPT sessions, or of Google queries, with at least one clean referral, so a session with several referrals counts once. The two denominators are different units of use.

\paragraph{Destination exclusions.} We exclude 147 destination domains that provide authentication, platform assets, infrastructure, advertising or tracking, and a few AI utilities. Examples are \texttt{microsoftonline.com}, \texttt{oaiusercontent.com}, \texttt{stripecdn.com}, \texttt{quillbot.com}, and \texttt{webpkgcache.com}. We also exclude search-engine domains and their subdomains, so the measure omits content and tools that search engines host themselves, such as \texttt{drive.google.com} and \texttt{docs.google.com}. Finally, we drop referrals from an intermediary, or from Gemini, to itself, and Google sign-in (OAuth) paths.

\paragraph{Attribution.} The referrer identifies the intermediary, and timing identifies the session or query. We assign each clean ChatGPT referral to the household's most recent conversation load, and through that load to its session (\autoref{oa:conv-cleaning}). A referral therefore belongs to a load if it occurs before the household's next conversation load. This window, from one load to the next, is separate from the one-hour rule that groups loads into sessions. For Google, a query is a foreground visit whose URL path ends in \path{/search}, and a clean referral belongs to a query if it occurs before the household's next search visit. A visit with no recorded referrer cannot be attributed, even when it follows use of an intermediary. The data show referrers and times; they do not show which link on a page the user clicked.

\paragraph{Checks.} Two checks support the measure (\autoref{tab:oa-referral-validation}). First, clean referrals come from the ChatGPT interface: 99.7\% have \texttt{chatgpt.com} as their referrer, against 69.1\% of all records referred from ChatGPT or another OpenAI host before filtering. Most of the records that the filters remove are data requests: 84.7\% of the unfiltered records have a JSON content type. Second, referrals from both intermediaries usually lead to a specific page. Among landing pages whose path can be classified, 75.6\% of ChatGPT referrals and 81.4\% of Google referrals reach a page below the homepage, and the median depth is one for both.

\begin{table}[!htbp]
  \centering
  \caption{Clean referrals come from the ChatGPT interface and mostly reach pages below the homepage.}
  \label{tab:oa-referral-validation}
  \footnotesize

  \begin{tabular}{lrr}
    \multicolumn{3}{l}{\textit{Panel A. Referrer host by filter stage (full sample, ten months)}}\\
    \toprule
    Filter stage & \texttt{chatgpt.com} & Other OpenAI (such as \texttt{auth.openai.com}) \\
    \midrule
    Before filtering (any content type) & 69.1\% & 30.9\% \\
    Clean referrals                     & 99.7\% &  0.3\% \\
    \bottomrule
  \end{tabular}

  \vspace{8pt}
  \begin{tabular}{lrrr}
    \multicolumn{4}{l}{\textit{Panel B. Landing-page depth}}\\
    \toprule
    Intermediary & Homepage & Below homepage & Median depth \\
    \midrule
    ChatGPT & 24.4\% & 75.6\% & 1.0 \\
    Google  & 18.6\% & 81.4\% & 1.0 \\
    \bottomrule
  \end{tabular}

  \exhibitnotes{Panel A reports the referrer host of ChatGPT-referred records before filtering and of clean referrals, in the full ten-month descriptive sample. Other OpenAI combines authentication, marketing, and remaining OpenAI subdomains. Panel B covers 5{,}318 ChatGPT referrals and 874{,}266 Google referrals with classifiable landing pages. Path depth counts the nonempty segments below the host: zero is a homepage and one or more is a deeper page.}
\end{table}

\paragraph{Message-level check.} Message sends are a finer unit than sessions, but they are well covered only in July~2025. In that month the full descriptive sample contains 184{,}030 logged-in sends by 6{,}056 users. The number of users exceeds the number in \autoref{tab:oa-endpoint-coverage}, which covers only the balanced panel. A recorded referral follows 1.3\% of logged-in sends and 1.3\% of anonymous sends, and it follows 1.0\% of logged-in sends within five minutes (\autoref{tab:oa-message-attribution}). These rates use a finer unit and a single month, so their levels differ from the session measure.

\begin{table}[!htbp]
  \centering
  \caption{Few observed July message sends are followed by a referral.}
  \label{tab:oa-message-attribution}
  \footnotesize

  \begin{tabular}{lrrr}
    \toprule
    Message sends (Jul 2025) & Sends & With referral & Share with referral \\
    \midrule
    Logged-in (\texttt{backend-api})  & 184{,}030 & 2{,}468 & 1.3\% \\
    \quad within 5 min                &           & 1{,}785 & 1.0\% \\
    Anonymous (\texttt{backend-anon}) &  31{,}320 &   411   & 1.3\% \\
    \bottomrule
  \end{tabular}

  \exhibitnotes{The unit is a July~2025 message-send request to the logged-in or anonymous \texttt{f/\allowbreak{}conversation} endpoint. Preparatory requests are excluded. The five-minute row restricts the logged-in numerator to referrals within that interval. July is the only month with substantial message-endpoint coverage; these estimates are supplementary to the ten-month session measure.}
\end{table}

\subsection{Within-household and context-adjusted comparisons}\label{oa:within-user-ref}

Main-paper \autoref{tab:referral-comparisons} compares referral incidence between ChatGPT and Google within households. Columns (1)--(3) compute incidence for each household, week, and intermediary. Columns (4) and (5) compute it separately for each surrounding context (\autoref{oa:context}), which gives more observations from the same households. All columns weight observations equally and cluster standard errors by household.

The context columns use a narrower sample. A ChatGPT session takes its own context label. A Google query takes the label of its search session, and its referrals are still attributed query by query. The sample keeps sessions near use of other search or AI services, and it drops search visits that fall outside any search session and contexts labeled solo, mixed, or other (\autoref{oa:context-cache}). Mean incidence in this sample is 0.088 for ChatGPT and 0.462 for Google. Because the sample and the level of aggregation differ, the coefficient in column (4), which has household and week fixed effects only, differs from the one in column (3). Surrounding browsing is a proxy for the task. Prompts and query text are unobserved, so the tasks brought to the two intermediaries can still differ within a context.

\subsection{Interpreting external benchmarks}\label{oa:ctr-bench}

Published traffic statistics for Google and ChatGPT use different units (\autoref{tab:oa-ctr-bench}). SparkToro and Datos count open-web clicks per Google query. Pew reports the share of result pages with a click on a traditional link. Momentic and Similarweb report outgoing visits per person, which combines how often people use a platform with how often each use produces a visit. Our measure is the share of sessions or queries with at least one referral.

\begin{table}[H]
  \centering
  \caption{Published traffic measures use different activity units and outcomes.}
  \label{tab:oa-ctr-bench}
  \footnotesize

  \begin{tabular}{p{4.7cm}p{6.0cm}r}
    \toprule
    Source & Metric & Value \\
    \midrule
    \multicolumn{3}{l}{\textit{Google}} \\
    Ours (Oct 2024--Jul 2025)               & Share of queries with a clean referral                              & 31.1\% \\
    SparkToro\,/\,Datos\tablefootnote{Rand Fishkin (SparkToro, on Datos clickstream data), \emph{2024 Zero-Click Search Study}, 1~July 2024, \url{https://sparktoro.com/blog/2024-zero-click-search-study-for-every-1000-us-google-searches-only-374-clicks-go-to-the-open-web-in-the-eu-its-360/} (accessed 2026-06-24).} (Jul 2024) & Open-web clicks per query (US)                  & 0.36 \\
    Pew Research Center\tablefootnote{A.~Chapekis and A.~Lieb, \emph{Google Users Are Less Likely to Click on Links When an AI Summary Appears in the Results}, Pew Research Center, 22~July 2025, \url{https://www.pewresearch.org/short-reads/2025/07/22/google-users-are-less-likely-to-click-on-links-when-an-ai-summary-appears-in-the-results/} (accessed 2026-06-24).} (Jul 2025) & Result pages without AI summary; any traditional-link click             & 15\% \\
    \midrule
    \multicolumn{3}{l}{\textit{ChatGPT}} \\
    Ours (Oct 2024--Jul 2025)                    & Share of ChatGPT sessions with a clean referral                                        & 5.4\% \\
    Ours (Jul 2025)             & Share of message sends followed by a referral                  & 1.3\% \\
    Momentic / Similarweb\tablefootnote{Tyler Einberger (Momentic), \emph{Google vs. ChatGPT}, May~2025, \url{https://momenticmarketing.com/blog/google-reach-chatgpt-click-throughs}, accompanying report pp.~12--13 and 22 (accessed 2026-09-21). The report defines the denominator as people (de-duplicated audience); the webpage also uses per-visit wording.} & Outgoing visits per person, March~2025 & 1.4 \\
    \bottomrule
  \end{tabular}
  \exhibitnotes{Our estimates use the full Comscore U.S. desktop sample from October~2024 through July~2025, except the July-only message check. External rows keep each source's population, period, and metric. SparkToro/Datos counts open-web clicks per query; Pew reports the share of result-page visits with a traditional-link click when no AI summary appears; Momentic reports outgoing visits per person.}\end{table}

\section{Surrounding browsing contexts}\label{oa:context}

Main-paper \autoref{sec:contexts} uses the websites a household visits just before and after a session to describe the circumstances of ChatGPT and Google use. This section defines those contexts. It then shows that the main patterns hold with longer windows, without sessions near other intermediaries, with a shorter Google session boundary, and with different thresholds for a dominant category. Website categories come from the classification in \autoref{oa:classification}.

\subsection{Sessions, windows, and category assignment}\label{oa:context-cache}

\paragraph{Sessions.} A context is attached to a ChatGPT session or to a Google search session. ChatGPT sessions follow the one-hour rule in \autoref{oa:conv-cleaning}. Google has no conversation identifier, so we group a household's queries into a session when successive queries are at most one hour apart, and we use thirty minutes as an alternative below. The two rules share a gap length and still define different units: a ChatGPT session follows one conversation, and a Google session follows a run of queries. Like ChatGPT sessions, Google sessions are built within calendar months. We use Google sessions only for the context analysis; the main referral comparison uses individual queries.

\paragraph{Nearby visits.} The context of a session is the set of foreground pages that the household loads in the fifteen minutes before the session starts and the fifteen minutes after it ends. A sensitivity check extends both windows to thirty minutes. A session counts as referring if it has at least one clean referral. For ChatGPT this follows the attribution rule in \autoref{oa:referral-def}; for Google sessions it counts referrals between the start of the session and the start of the next one.

\paragraph{Dominant category.} We label each context by the category of the websites in it. A nearby visit is classifiable if its destination is one of the 3{,}245 domains classified with confidence of at least 0.90, and if neither its destination nor its referrer is the session's own intermediary, a search engine, or an LLM platform. A category is dominant when it accounts for at least half of a session's classifiable visits (threshold $\tau=0.50$). A session with classifiable visits and no dominant category is a \emph{mixed context}, and a session with no classifiable visits is \emph{solo}. The context-adjusted regressions drop solo, mixed, and other contexts (\autoref{oa:within-user-ref}).

\subsection{Longer browsing windows}\label{oa:context-window}

\begin{figure}[!htbp]
  \centering
  \caption{Differences in surrounding browsing composition persist with the longer window.}
  \label{fig:oa-context-30min-composition}

  \begin{subfigure}{0.49\linewidth}
    \centering
    \includegraphics[width=\linewidth]{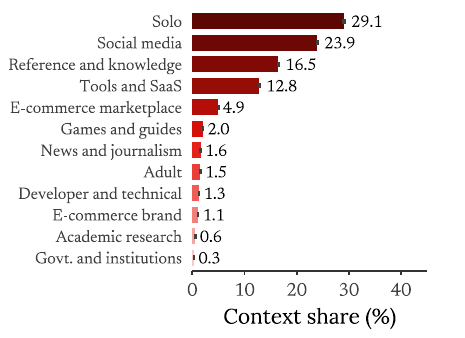}
    \caption{ChatGPT.}
  \end{subfigure}\hfill
  \begin{subfigure}{0.49\linewidth}
    \centering
    \includegraphics[width=\linewidth]{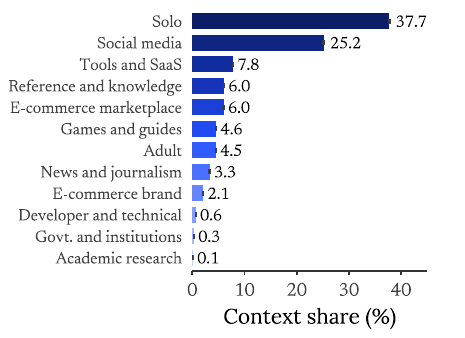}
    \caption{Google.}
  \end{subfigure}

  \exhibitnotes{The panels show dominant content categories using thirty-minute windows before and after each session, keeping all sessions. Both intermediaries use a one-hour session gap. Shares use all sessions as the denominator; mixed and other categories are omitted from the display. The sample is all observed active desktop households over the ten-month window.}
\end{figure}

\begin{figure}[!htbp]
  \centering
  \caption{ChatGPT referral incidence continues to vary across thirty-minute contexts.}
  \label{fig:oa-context-30min-rr}

  \includegraphics[width=3.10in]{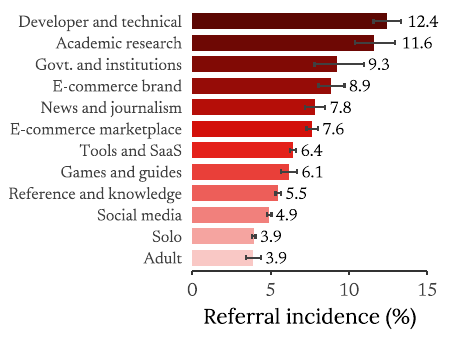}

  \exhibitnotes{Bars report the share of ChatGPT sessions with at least one clean referral, conditional on the dominant surrounding category in thirty-minute windows. All sessions are kept, including those near use of other intermediaries. Error bars are 95\% confidence intervals. Solo sessions have no classifiable surrounding visits.}
\end{figure}

The differences between ChatGPT and Google persist when each window is extended from fifteen to thirty minutes, which adds more distant browsing. The composition of contexts still differs between the intermediaries (\autoref{fig:oa-context-30min-composition}), and ChatGPT's referral incidence still varies across contexts (\autoref{fig:oa-context-30min-rr}).

\subsection{Excluding nearby use of other intermediaries}\label{oa:context-pure-vs-mixed}

People often use ChatGPT and a search engine close together. Within thirty minutes before or after a ChatGPT session, 79\% of sessions have a visit to or from Google, Bing, or Yahoo. Only 3\% of Google sessions have nearby activity on an LLM platform. The main comparison keeps all sessions. A restricted comparison drops sessions with this nearby activity, which leaves 85{,}316 of 409{,}133 ChatGPT sessions.

\begin{figure}[!htbp]
  \centering
  \caption{Excluding sessions near other intermediaries changes context shares more than referral-incidence patterns.}
  \label{fig:oa-context-pure-mixed-comp}
  \begin{subfigure}{0.49\linewidth}
    \centering
    \includegraphics[width=\linewidth]{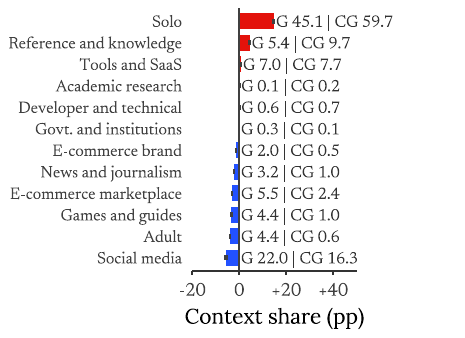}
    \caption{ChatGPT minus Google context shares.}
  \end{subfigure}\hfill
  \begin{subfigure}{0.49\linewidth}
    \centering
    \includegraphics[width=\linewidth]{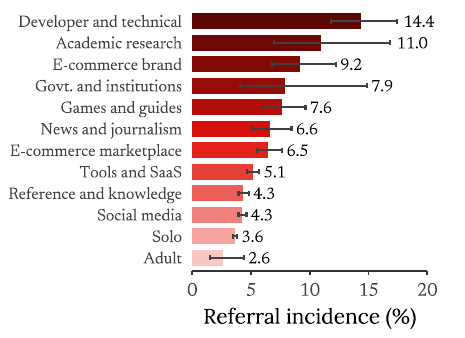}
    \caption{ChatGPT referral incidence by context.}
  \end{subfigure}
  \exhibitnotes{Both panels exclude sessions with other-intermediary activity within thirty minutes before or after the session. Displayed contexts use fifteen-minute windows and one-hour session boundaries. Panel (a) reports ChatGPT minus Google category shares after excluding mixed and other contexts; panel (b) reports ChatGPT session referral incidence within each displayed category. Error bars are 95\% confidence intervals. Main-paper \autoref{fig:context} shows the corresponding estimates keeping all sessions.}
\end{figure}

The restriction changes ChatGPT's context shares mainly by making more of its sessions solo (\autoref{fig:oa-context-pure-mixed-comp}, panel (a)). Among the displayed shares, which exclude mixed and other contexts, ChatGPT's solo share rises from 41.2\% to 59.7\%, and its lead over Google in reference/knowledge and tools/SaaS contexts narrows. In the restricted sample, reference/knowledge accounts for 9.7\% of ChatGPT sessions and 5.4\% of Google sessions, and tools/SaaS for 7.7\% and 7.0\%. Sessions that do have surrounding browsing look much the same in both samples. Among sessions that are not solo, reference/knowledge accounts for about a quarter of ChatGPT sessions and tools/SaaS for about a fifth in both samples, against about one in ten and one in eight for Google.

Referral incidence within contexts also keeps its pattern in the much smaller restricted sample (panel (b), compared with main-paper \autoref{fig:context}). Developer/technical contexts have high incidence in both samples, 13.6\% with all sessions and 14.4\% in the restricted sample, and adult contexts have low incidence, 3.9\% and 2.6\%.

\subsection{Google session boundaries}\label{oa:context-gap-pair}

\begin{figure}[!htbp]
  \centering
  \caption{The context-composition comparison is similar with a thirty-minute Google session boundary.}
  \label{fig:oa-context-gap-comp}
  \includegraphics[width=3.10in]{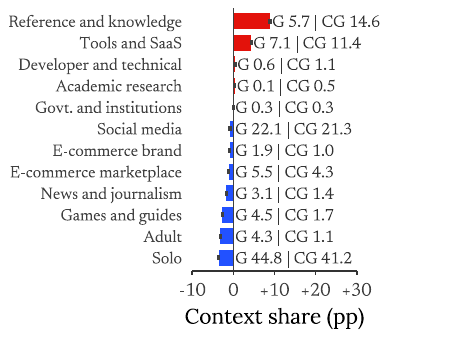}
  \exhibitnotes{Bars show ChatGPT minus Google category-share differences with fifteen-minute browsing windows, keeping sessions with nearby use of other intermediaries and excluding mixed and other contexts from the shares. Google's maximum inter-query gap is thirty minutes; ChatGPT sessions use the one-hour rule. Error bars are 95\% confidence intervals. Main-paper \autoref{fig:context}, panel (a), reports the one-hour Google boundary.}
\end{figure}

The comparison of context shares is similar when Google's session gap is thirty minutes instead of one hour (\autoref{fig:oa-context-gap-comp}). The shorter gap splits some multi-query sessions and changes the visits that surround them. ChatGPT sessions and their contexts are the same as in the main comparison.

\subsection{Thresholds for a dominant category}\label{oa:context-tau}

Referral incidence within named categories barely moves when the share required for a dominant category varies from 0.30 to 0.60 (\autoref{tab:oa-dominance-sweep}, with thirty-minute windows). ChatGPT's incidence changes by less than half a percentage point in every named category, and Google's session-level incidence by less than one point. A higher threshold labels more sessions mixed, so the mixed row varies more. The solo row is constant because solo sessions have no classifiable visits.

\begin{table}[!htbp]
  \centering
  \caption{Referral incidence within named categories is stable across dominance thresholds.}
  \label{tab:oa-dominance-sweep}
  \footnotesize

  \begin{tabular}{lrrrr}
    \toprule
    & \multicolumn{4}{c}{Referral incidence (\%), by dominance threshold $\tau$} \\
    \cmidrule(lr){2-5}
    Surrounding context & 0.3 & 0.4 & 0.5 & 0.6 \\
    \midrule
    \multicolumn{5}{l}{\emph{Panel A: ChatGPT sessions}} \\
    \midrule
    Developer/technical & 12.5 & 12.5 & 12.4 & 12.1 \\
    Academic research & 11.9 & 11.8 & 11.6 & 11.8 \\
    Government/institutional & 9.2 & 9.3 & 9.3 & 9.3 \\
    E-commerce brand & 9.0 & 9.0 & 8.9 & 8.7 \\
    E-commerce marketplace & 7.7 & 7.7 & 7.6 & 7.6 \\
    News/journalism & 7.9 & 7.9 & 7.8 & 7.6 \\
    Tools/SaaS & 6.5 & 6.5 & 6.4 & 6.2 \\
    Games and guides & 6.2 & 6.2 & 6.1 & 6.2 \\
    Other & 6.1 & 6.0 & 6.0 & 5.9 \\
    Reference/knowledge & 5.5 & 5.5 & 5.5 & 5.4 \\
    Social media & 5.0 & 5.0 & 4.9 & 4.7 \\
    Adult & 4.0 & 3.9 & 3.9 & 3.9 \\
    \addlinespace
    Solo & 3.9 & 3.9 & 3.9 & 3.9 \\
    Mixed & 10.4 & 7.8 & 8.3 & 7.4 \\
    \midrule
    \multicolumn{5}{l}{\emph{Panel B: Google sessions}} \\
    \midrule
    Developer/technical & 78.4 & 78.4 & 78.4 & 78.3 \\
    Academic research & 63.2 & 63.1 & 63.0 & 62.7 \\
    Government/institutional & 72.4 & 72.3 & 72.3 & 72.0 \\
    E-commerce brand & 75.8 & 75.8 & 75.8 & 75.7 \\
    E-commerce marketplace & 69.9 & 69.9 & 69.8 & 69.5 \\
    News/journalism & 64.9 & 64.9 & 64.8 & 64.5 \\
    Tools/SaaS & 68.7 & 68.7 & 68.6 & 68.4 \\
    Games and guides & 78.9 & 78.9 & 78.9 & 78.9 \\
    Other & 70.1 & 70.0 & 69.9 & 69.6 \\
    Reference/knowledge & 67.9 & 67.9 & 67.8 & 67.4 \\
    Social media & 66.5 & 66.4 & 66.3 & 65.8 \\
    Adult & 83.3 & 83.3 & 83.3 & 83.4 \\
    \addlinespace
    Solo & 55.3 & 55.3 & 55.3 & 55.3 \\
    Mixed & 72.5 & 73.6 & 74.6 & 72.8 \\
    \bottomrule
  \end{tabular}
  \exhibitnotes{Cells are percentages of each intermediary's sessions with at least one clean referral, using all sessions, one-hour session boundaries, and thirty-minute windows. The columns vary the minimum share of classifiable surrounding visits needed for a dominant category. Solo sessions have no classifiable surrounding visits; mixed contexts have such visits but no category meeting the threshold. Google denominators here are sessions; the main comparison uses individual queries.}
\end{table}

\section{Destination classification and composition}\label{oa:classification}

Main-paper \autoref{sec:selective} compares the kinds of websites that receive ChatGPT and Google referrals. The comparison requires a category for each destination. This section describes the domains we classify, the categories, how a language model assigned them, and how its labels compare with an author's independent coding.

\subsection{Domains and categories}\label{oa:taxonomy}

We classify 4{,}266 domains: the union of the 2{,}500 domains that receive the most referrals from LLM platforms and the 2{,}500 that receive the most from search engines. Each list covers about 71\% of its source's referrals. Raising the cutoff to 10{,}000 domains per list would cover 94\% of LLM referrals and 81\% of search referrals, and would require classifying 17{,}545 domains (\autoref{tab:oa-domain-coverage}).

\begin{table}[!htbp]
  \centering
  \caption{Broader rank cutoffs add coverage at the cost of a much larger domain universe.}
  \label{tab:oa-domain-coverage}
  \footnotesize

  \begin{tabular}{rrrrrrr}
    \toprule
    & \multicolumn{3}{c}{Domains} & \multicolumn{3}{c}{Referral coverage} \\
    \cmidrule(lr){2-4}\cmidrule(lr){5-7}
    $K$ & LLM list only & Both lists & Union & LLM & Search & LLM, union \\
    \midrule
    500   & 344    & 156   & 844    & 0.515 & 0.578 & 0.530 \\
    1{,}000 & 697    & 303   & 1{,}697  & 0.593 & 0.635 & 0.608 \\
    2{,}000 & 1{,}399  & 601   & 3{,}399  & 0.680 & 0.690 & 0.696 \\
    2{,}500 (main) & 1{,}766  & 734   & 4{,}266  & 0.711 & 0.708 & 0.726 \\
    3{,}000 & 2{,}154  & 846   & 5{,}154  & 0.735 & 0.722 & 0.751 \\
    5{,}000 & 3{,}658  & 1{,}342 & 8{,}658  & 0.821 & 0.760 & 0.834 \\
    10{,}000 & 7{,}545  & 2{,}455 & 17{,}545 & 0.941 & 0.807 & 0.948 \\
    \bottomrule
  \end{tabular}
  \exhibitnotes{$K$ is applied separately to the LLM- and search-referral rankings in the full ten-month descriptive sample. The middle columns count domains in only the LLM list, in both lists, and in their union. Referral coverage is the share of each source\textquotesingle s referrals that reach its own top-$K$ list; the last column is the share of LLM referrals that reach the union. The classification universe uses $K=2{,}500$.}
\end{table}

Each domain receives two labels, one for its content and one for how it earns revenue (\autoref{tab:oa-codebook}). A label describes the domain's primary activity and its visible business model. Three choices matter for reading the results. Education services and health-reference websites count as reference/knowledge, and wikis devoted to a single game count as games and guides. Social media includes video and audio platforms, among them professional streaming libraries. The category other holds general-purpose AI assistants and AI-only generators along with websites the classifier could not place. \autoref{tab:oa-classification-dist} reports how many domains fall in each category.

\begin{table}[!htbp]
  \centering
  \caption{Each domain receives a content category and a monetization category.}
  \label{tab:oa-codebook}
  \footnotesize
  \begin{tabular}{@{}>{\raggedright\arraybackslash}p{0.27\linewidth} >{\raggedright\arraybackslash}p{0.69\linewidth}@{}}
    \toprule
    Category & Definition and examples \\
    \midrule
    \multicolumn{2}{@{}l}{\emph{Content type}} \\
    Reference/knowledge & Encyclopedic, how-to, dictionary, and tutorial content, including education services and health-reference websites (\texttt{wikipedia.org}, \texttt{mayoclinic.org}) \\
    News/journalism & Original reporting, current news, and opinion (\texttt{nytimes.com}, \texttt{reuters.com}) \\
    Academic research & Peer-reviewed journals, preprint servers, research organizations, and academic databases (\texttt{arxiv.org}, \texttt{nature.com}) \\
    Social media & Social, question-and-answer, and discussion platforms, and video or audio platforms, including professional streaming libraries (\texttt{reddit.com}, \texttt{youtube.com}, \texttt{netflix.com}) \\
    E-commerce brand & Manufacturers and direct-to-consumer websites selling their own goods or services (\texttt{nike.com}, \texttt{apple.com}) \\
    E-commerce marketplace & Multi-seller marketplaces and broad-line retailers (\texttt{amazon.com}, \texttt{ebay.com}) \\
    Tools/SaaS & Productivity tools, web applications, and utilities (\texttt{canva.com}, \texttt{notion.so}) \\
    Developer/technical & Code repositories, package registries, and developer documentation (\texttt{github.com}, \texttt{npmjs.com}) \\
    Government/institutional & Government websites, public-service portals, and international institutions (\texttt{europa.eu}, \texttt{who.int}) \\
    Games and guides & Gaming websites, mod communities, fan-fiction archives, and wikis devoted to a single game (\texttt{steamcommunity.com}, \texttt{fandom.com}) \\
    Adult & Explicit-content websites \\
    Other & General-purpose AI assistants and AI-only generators (\texttt{chatgpt.com}, \texttt{claude.ai}), and websites the classifier could not place \\
    \midrule
    \multicolumn{2}{@{}l}{\emph{Monetization}} \\
    Ad-supported & Free access, with advertising as the main visible source of revenue (\texttt{cnn.com}, \texttt{weather.com}) \\
    Subscription/paywall & A recurring fee for access (\texttt{wsj.com}, \texttt{netflix.com}) \\
    Freemium SaaS & A free tier with a paid upgrade (\texttt{dropbox.com}, \texttt{grammarly.com}) \\
    Transactional & Revenue from on-site sales of goods or services (\texttt{amazon.com}, \texttt{expedia.com}) \\
    Non-profit/public & Funded by a foundation, a government, or a non-profit institution (\texttt{wikipedia.org}, \texttt{mayoclinic.org}) \\
    Mixed/other & Two or more substantial revenue sources (\texttt{nytimes.com}, \texttt{twitch.tv}) \\
    \bottomrule
  \end{tabular}
  \exhibitnotes{Definitions and examples are condensed from the instructions given to the classifier.}
\end{table}

\begin{table}[!htbp]
  \centering
  \caption{The domain universe spans multiple content and business-model categories.}
  \label{tab:oa-classification-dist}
  \footnotesize

  \begin{tabular}{lr@{\hspace{0.5em}}r@{\hspace{2em}}lr@{\hspace{0.5em}}r}
    \toprule
    \multicolumn{3}{l}{\textbf{Content type}} & \multicolumn{3}{l}{\textbf{Monetization}} \\
    Category & $N$ & \% & Category & $N$ & \% \\
    \midrule
    Reference/knowledge     & 803 & 18.8 & Ad-supported        & 1{,}708 & 40.0 \\
    Tools/SaaS              & 686 & 16.1 & Transactional        & 890     & 20.9 \\
    Other                    & 490 & 11.5 & Non-profit/public  & 680     & 15.9 \\
    E-commerce brand         & 425 & 10.0 & Freemium SaaS       & 578     & 13.5 \\
    Games and guides    & 348 & 8.2  & Subscription/paywall & 248    & 5.8 \\
    Government/institutional & 345 & 8.1 & Mixed/other         & 137     & 3.2 \\
    News/journalism         & 327 & 7.7  & Unknown              & 25      & 0.6 \\
    E-commerce marketplace   & 259 & 6.1  & & & \\
    Adult                    & 250 & 5.9  & & & \\
    Social media            & 166 & 3.9  & & & \\
    Academic research       &  82 & 1.9  & & & \\
    Developer/technical     &  80 & 1.9  & & & \\
    Unknown                  &   5 & 0.1  & & & \\
    \bottomrule
  \end{tabular}
  \exhibitnotes{Counts and percentages describe all 4{,}266 domains in the union of the two top-2{,}500 lists. The two dimensions are classified separately. Other holds AI assistants and generators along with websites the classifier could not place; unknown marks domains without a valid label. Shares count domains; referral-weighted shares appear in main-paper \autoref{fig:composition}.}
\end{table}

\subsection{Assigning labels}\label{oa:labeling}

GPT-4o assigned the labels. Its instructions gave the definitions and examples in \autoref{tab:oa-codebook} and asked, for each domain, for both labels, a short rationale, and a confidence score between zero and one. The model ran at temperature zero and could search the web when it did not recognize a website. The main content analyses use the 3{,}245 domains with confidence of at least 0.90.

We then override the model for two groups of domains: \texttt{.gov} domains are assigned to reference/knowledge and \texttt{.edu} domains to academic research, both with non-profit/public monetization. This recoding applies wherever categories are used, in destination composition, surrounding contexts, and the category outcomes in \autoref{oa:category-outcomes}. After the recoding, the government/institutional category holds public institutions outside \texttt{.gov}. The counts in \autoref{tab:oa-classification-dist} and the validation below describe the model's labels before the recoding. In the composition analysis, content shares exclude the other and unknown labels, and monetization shares exclude unknown and keep mixed/other.

\subsection{Validation against blind human coding}\label{oa:classification-validation}

One author coded 100 domains with the same codebook, without seeing the model's labels. The model and the author agree on 79\% of domains for content type and 87\% for monetization. Cohen's $\kappa$, which corrects for chance agreement, is 0.76 and 0.82 (\autoref{tab:oa-val-agreement}; intervals use 5{,}000 bootstrap resamples).

The sample was drawn at random within three strata of referral rank: forty domains from ranks 1--200, thirty from 201--1{,}000, and thirty from 1{,}001 onward. It spans ranks 10 to 4{,}082 and about 7.5\% of referral volume. Stratifying puts domains beyond the largest websites into the sample, and design weights correct for the unequal sampling rates.

\begin{table}[!htbp]
  \centering
  \caption{Model and human labels agree on most sampled domains.}
  \label{tab:oa-val-agreement}
  \footnotesize
  \setlength{\tabcolsep}{4pt}

  \begin{tabular}{llrrrrrl}
    \toprule
    Subset & Dimension & $n$ & Categories & Exact & Weighted & $\kappa$ & 95\% CI \\
    \midrule
    Overall              & Content type & 100 & 11 & 0.79 & 0.81 & 0.76 & [0.67, 0.85] \\
    Overall              & Monetization & 100 & 6  & 0.87 & 0.91 & 0.82 & [0.72, 0.90] \\
    Without model's ``other'' & Content type & 88 & 11 & 0.84 & 0.84 & 0.82 & [0.72, 0.90] \\
    Without model's ``other'' & Monetization & 88 & 6  & 0.88 & 0.90 & 0.82 & [0.72, 0.92] \\
    \bottomrule
  \end{tabular}
  \exhibitnotes{The reference is one blind human coder on the stratified 100-domain sample. Exact and weighted agreement are proportions; the weighted measure weights domains by referral volume within this sample. Cohen\textquotesingle s $\kappa$ adjusts for chance agreement; the interval is a 95\% percentile-bootstrap interval for $\kappa$ from 5{,}000 resamples. The conditional rows exclude twelve model assignments to other, including AI services as well as sites the model could not place, and so describe agreement among the remaining 88 domains.}
\end{table}

Thirty domains disagree on at least one label: seventeen on content only, nine on monetization only, and four on both. Part of the content disagreement comes from the category other, which the model uses for twelve domains and the author for six, five of them the same. Without the model's twelve other assignments, content agreement is 84\%. Agreement is weakest for reference/knowledge and news/journalism, where the model's precision is 0.55 and 0.56, and for mixed/other monetization, where recall is 0.29 (\autoref{tab:oa-val-percat}). Several categories have few validation domains, so the category-level figures are imprecise.

\begin{table}[!htbp]
  \centering
  \caption{Agreement varies across categories.}
  \label{tab:oa-val-percat}
  \footnotesize

  \begin{tabular}{lrrrrr}
    \toprule
    Category & Human count & Model count & Precision & Recall & $F_1$ \\
    \midrule
    \multicolumn{6}{l}{\textbf{Content type}} \\
    Games and guides    & 25 & 17 & 0.94 & 0.64 & 0.76 \\
    Tools/SaaS              & 17 & 13 & 0.92 & 0.71 & 0.80 \\
    E-commerce brand         & 13 & 14 & 0.93 & 1.00 & 0.96 \\
    Adult                    & 12 & 10 & 1.00 & 0.83 & 0.91 \\
    Reference/knowledge     &  7 & 11 & 0.55 & 0.86 & 0.67 \\
    Other                    &  6 & 12 & 0.42 & 0.83 & 0.56 \\
    E-commerce marketplace   &  5 &  6 & 0.83 & 1.00 & 0.91 \\
    News/journalism         &  5 &  9 & 0.56 & 1.00 & 0.71 \\
    Social media            &  5 &  4 & 0.75 & 0.60 & 0.67 \\
    Government/institutional &  4 &  3 & 1.00 & 0.75 & 0.86 \\
    Academic research       &  1 &  1 & 1.00 & 1.00 & 1.00 \\
    \midrule
    \multicolumn{6}{l}{\textbf{Monetization}} \\
    Ad-supported            & 43 & 48 & 0.85 & 0.95 & 0.90 \\
    Transactional            & 25 & 23 & 1.00 & 0.92 & 0.96 \\
    Freemium SaaS           & 11 & 12 & 0.75 & 0.82 & 0.78 \\
    Non-profit/public      &  9 &  7 & 1.00 & 0.78 & 0.88 \\
    Mixed/other             &  7 &  3 & 0.67 & 0.29 & 0.40 \\
    Subscription/paywall    &  5 &  7 & 0.71 & 1.00 & 0.83 \\
    \bottomrule
  \end{tabular}
  \exhibitnotes{Precision, recall, and $F_1$ compare model assignments with the human reference in the full 100-domain validation sample. Human and model counts are the numbers of validation domains that the human coder and the model assign to each category. Small category counts, including one academic-research domain, limit the precision of category-specific assessments.}
\end{table}

Weighting the sample back to all 4{,}266 domains gives per-domain agreement of 0.77 for content and 0.79 for monetization. When domains are weighted by their referral traffic, agreement is 0.79 and 0.88 (\autoref{tab:oa-val-strata}).

\begin{table}[!htbp]
  \centering
  \caption{Design weighting extends the validation beyond the sampled domains.}
  \label{tab:oa-val-strata}
  \footnotesize

  \resizebox{\linewidth}{!}{%
\begin{tabular}{lrrrrr}
    \toprule
    & & \multicolumn{2}{c}{Content type} & \multicolumn{2}{c}{Monetization} \\
    \cmidrule(lr){3-4}\cmidrule(lr){5-6}
    \multicolumn{2}{l}{\textbf{Panel A: by referral-rank stratum}} & Exact & $\kappa$ & Exact & $\kappa$ \\
    \midrule
    Head (rank 1--200)        & $n{=}40$ & 0.80 & 0.76 & 0.95 & 0.92 \\
    Middle (rank 201--1{,}000) & $n{=}30$ & 0.80 & 0.77 & 0.87 & 0.80 \\
    Tail (rank 1{,}001$+$)     & $n{=}30$ & 0.77 & 0.73 & 0.77 & 0.71 \\
    \midrule
    \multicolumn{2}{l}{\textbf{Panel B: population estimates}} & Content type & & Monetization & Coverage \\
    \midrule
    \multicolumn{2}{l}{Design-weighted, per-domain (Horvitz--Thompson)} & 0.77 & & 0.79 & 100\% \\
    \multicolumn{2}{l}{Post-stratified, traffic-weighted}               & 0.79 & & 0.88 & approximately 99\% \\
    \bottomrule
  \end{tabular}
}
  \exhibitnotes{Panel A reports agreement within each referral-rank stratum. Panel B weights sampled domains by inverse sampling fractions for the per-domain estimates and uses universe referral shares for the post-stratified traffic-weighted estimates. Coverage reports the portion of the target domain or traffic universe represented. The validation sample has forty head, thirty middle, and thirty tail domains.}
\end{table}

\subsection{Confidence scores}\label{oa:confidence-calibration}

The model's confidence score separates more reliable from less reliable content labels. Among validation domains with confidence of at least 0.90, the content label matches the author's on 83.5\% ($\kappa=0.813$); below 0.90 it matches on 61.9\% ($\kappa=0.525$). Monetization agreement does not rise with confidence (\autoref{tab:oa-confidence-agreement}). We therefore use the score to rank content labels by reliability, and we do not read it as the probability that a label is correct. Of the 4{,}266 domains, 76.1\% reach the 0.90 threshold and 94.7\% reach 0.60.

\begin{table}[!htbp]
  \centering
  \caption{Higher confidence predicts stronger content agreement, but not stronger monetization agreement.}
  \label{tab:oa-confidence-agreement}
  \footnotesize

  \begin{tabular}{lrrrrr}
    \toprule
    & & \multicolumn{2}{c}{Content type} & \multicolumn{2}{c}{Monetization} \\
    \cmidrule(lr){3-4}\cmidrule(lr){5-6}
    Subset & $n$ & Exact & $\kappa$ & Exact & $\kappa$ \\
    \midrule
    Full sample          & 100 & 0.790 & 0.761 & 0.870 & 0.817 \\
    Confidence $\geq 0.90$  &  79 & 0.835 & 0.813 & 0.861 & 0.813 \\
    Confidence $< 0.90$     &  21 & 0.619 & 0.525 & 0.905 & 0.768 \\
    \bottomrule
  \end{tabular}
  \exhibitnotes{The 100-domain validation sample is split at confidence 0.90. Exact is the model--human agreement share; $\kappa$ is chance-corrected agreement. These conditional comparisons are descriptive, with different sample sizes, and assess how well confidence separates agreeing from disagreeing labels.}
\end{table}

\subsection{A lower confidence threshold}\label{oa:composition-confidence}

\begin{figure}[!htbp]
  \centering
  \caption{The main composition differences persist at the lower confidence threshold.}
  \label{fig:oa-composition-robust}

  \begin{subfigure}{0.49\linewidth}
    \includegraphics[width=\linewidth]{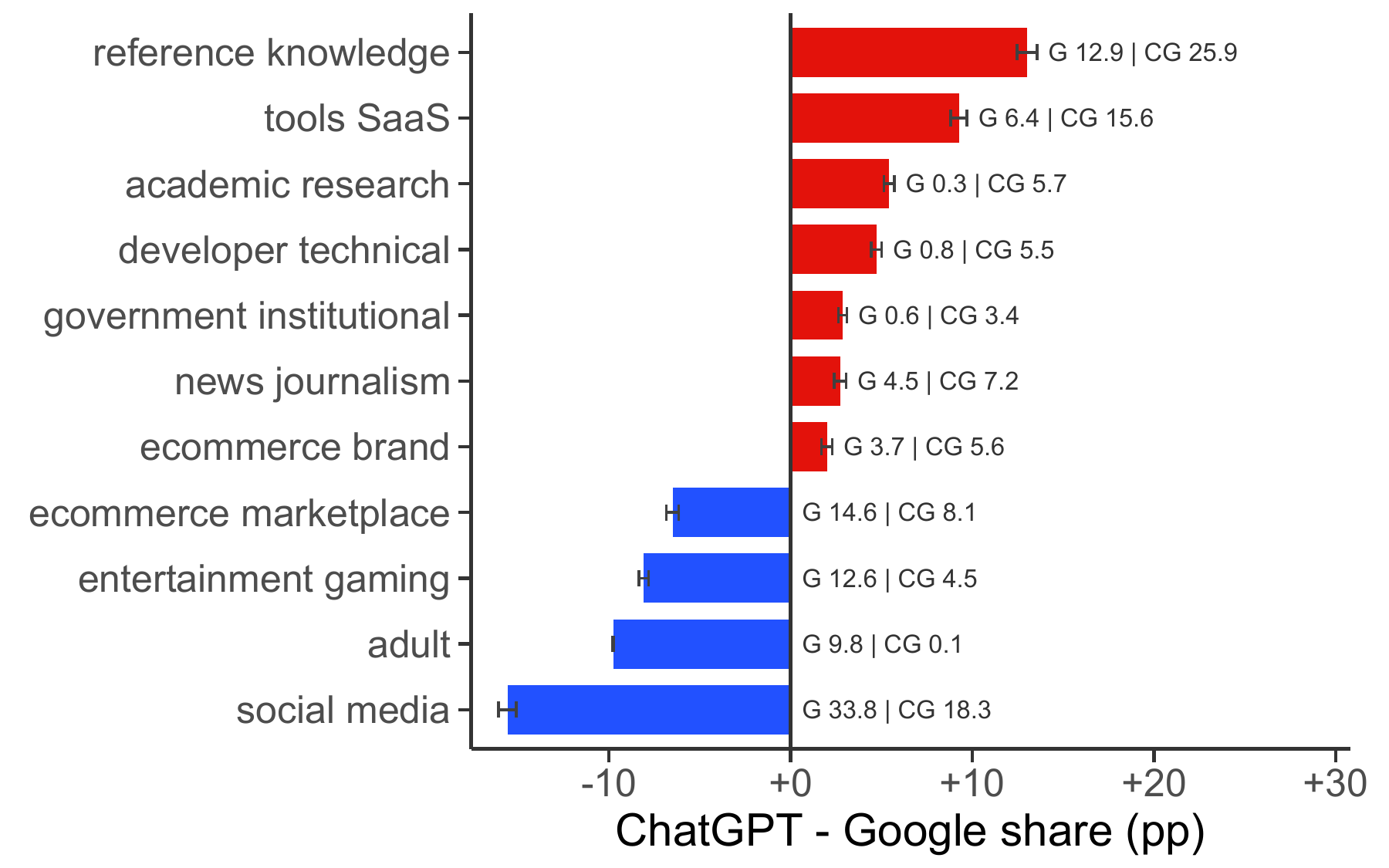}
    \caption{Content type.}
  \end{subfigure}\hfill
  \begin{subfigure}{0.49\linewidth}
    \includegraphics[width=\linewidth]{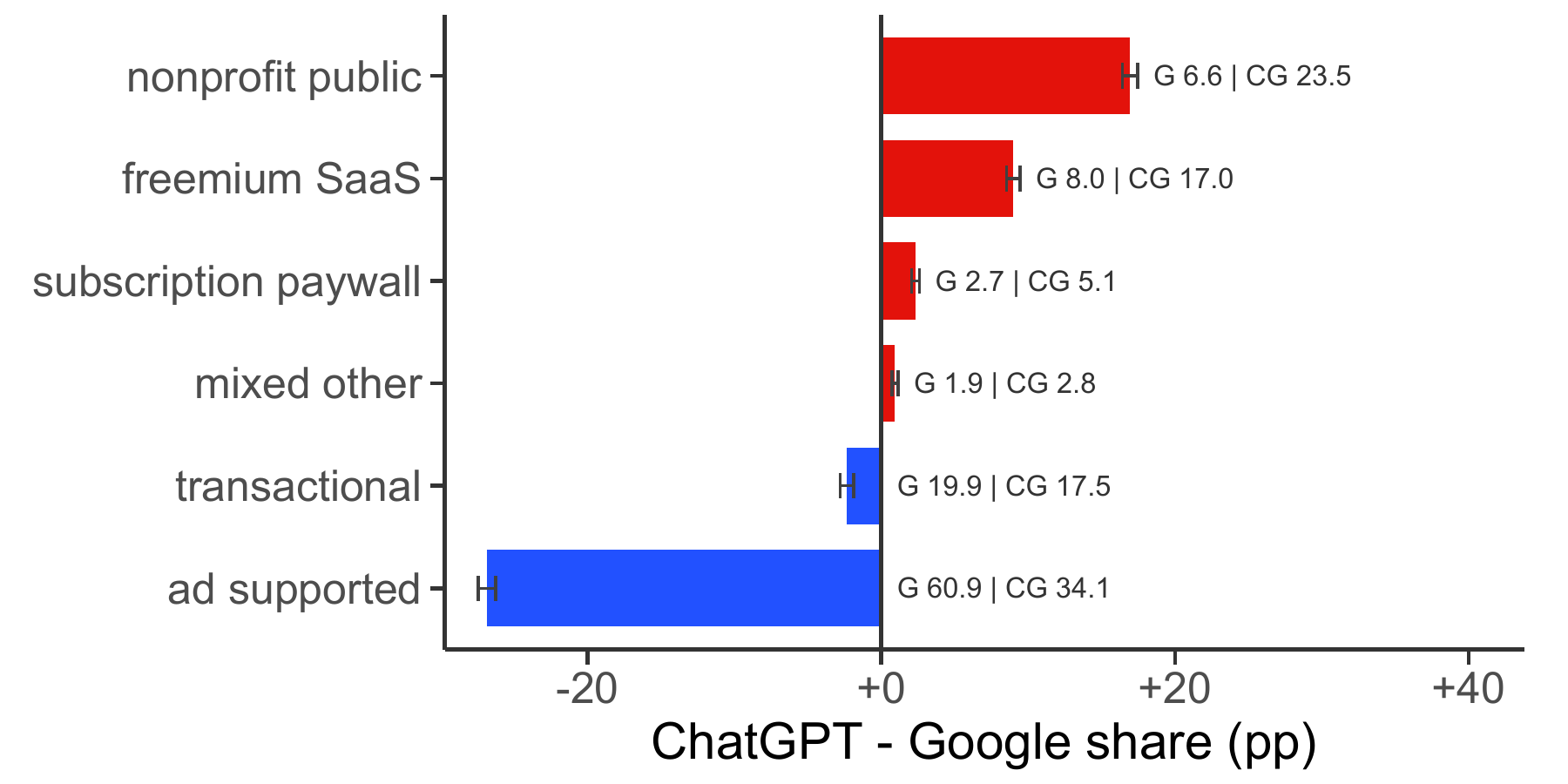}
    \caption{Monetization.}
  \end{subfigure}

  \exhibitnotes{Bars report ChatGPT\textquotesingle s referral share minus Google\textquotesingle s by content type (a) and monetization (b) after relaxing the classifier-confidence threshold from 0.90 to 0.60. The domain universe is the union of the two top-2{,}500 referral lists. Shares use domains meeting the 0.60 threshold after the recoding in \autoref{oa:labeling}. Content shares exclude other and unknown; monetization shares exclude unknown and retain mixed/other. Error bars are pointwise 95\% normal intervals for differences in proportions.}
\end{figure}

The composition differences hold when the threshold is lowered from 0.90 to 0.60, which adds about 790 domains. ChatGPT still sends larger shares of its referrals to reference, academic, developer, and tool websites, and a smaller share to ad-supported websites (\autoref{fig:oa-composition-robust}).

\section{Destination concentration and supplementary characteristics}\label{oa:concentration}

Main-paper \autoref{sec:concentration} compares how concentrated ChatGPT and Google referrals are, across all referrals and within household-weeks. This section defines the index, varies the set of destinations and the minimum number of referrals, and relates referral shares to website prominence.

\subsection{Normalized concentration}\label{oa:hhi-def}

Let $s_i$ be the share of an intermediary's referrals that go to destination $i$, among $n$ observed destinations. The Herfindahl index is $H=\sum_i s_i^2$. Its smallest possible value, $1/n$, is higher when fewer destinations are observed, so we use the normalized index
\begin{equation}
H^*=\frac{H-1/n}{1-1/n},\qquad n\geq2,
\label{eq:oa-normhhi}
\end{equation}
which is zero when all destinations have equal shares and approaches one when a single destination receives nearly all referrals. The index is undefined for a single destination, and we drop those observations.

\subsection{Restricting the set of destinations}\label{oa:hhi}

Google's referrals are more concentrated than ChatGPT's under every definition of the destination set in \autoref{tab:oa-hhi}. The table restricts destinations by traffic percentile, rank, cumulative traffic share, a minimum number of referrals from both intermediaries, and membership in the 4{,}266 classified domains, and it recomputes shares within each set. The ratio of Google's normalized index to ChatGPT's is 3.47 across all domains. It is smallest, 1.87, among each intermediary's top 100 destinations, and largest, 4.86, among destinations with at least five referrals from each intermediary. The direction never changes; the size of the ratio depends on which destinations enter.

The classified domains give a ratio of 2.68. That set combines the domains most referred by LLM platforms and by search engines (\autoref{oa:classification}). Main-paper \autoref{tab:hhi} uses a slightly different set, the union of ChatGPT's and Google's own 2{,}500 most-referred domains (4{,}026 domains), and reports a ratio of 2.94.

\begin{table}[!htbp]
  \centering
  \caption{Google referrals remain more concentrated under every destination restriction.}
  \label{tab:oa-hhi}
  \scriptsize

  \begin{tabular}{llrrrr}
    \toprule
    Rule & Destinations included & ChatGPT $H$ & Google $H$ & Ratio & Normalized ratio \\
    \midrule
    Percentile  & Top 1\% of each intermediary (union)             & 0.0120 & 0.0273 & 2.27 & 2.30 \\
    Percentile  & Top 5\% of each intermediary (union)             & 0.0085 & 0.0230 & 2.70 & 2.72 \\
    Percentile  & Top 10\% of each intermediary (union)            & 0.0076 & 0.0218 & 2.87 & 2.89 \\
    Percentile  & Top 25\% of each intermediary (union)            & 0.0066 & 0.0206 & 3.10 & 3.12 \\
    Percentile  & Top 50\% of each intermediary (union)            & 0.0061 & 0.0198 & 3.27 & 3.29 \\
    \midrule
    Rank        & Top 100 of each intermediary (union)             & 0.0425 & 0.0731 & 1.72 & 1.87 \\
    Rank        & Top 500 of each intermediary (union)             & 0.0198 & 0.0469 & 2.37 & 2.47 \\
    Rank        & Top 1{,}000 of each intermediary (union)          & 0.0155 & 0.0402 & 2.59 & 2.67 \\
    Rank        & Top 5{,}000 of each intermediary (union)          & 0.0095 & 0.0298 & 3.13 & 3.17 \\
    Rank        & Top 10{,}000 of each intermediary (union)         & 0.0079 & 0.0271 & 3.44 & 3.47 \\
    \midrule
    Cumulative share & 50\% of each intermediary\textquotesingle s traffic (union)          & 0.0218 & 0.0632 & 2.89 & 3.09 \\
    Cumulative share & 80\% of each intermediary\textquotesingle s traffic (union)          & 0.0085 & 0.0290 & 3.42 & 3.46 \\
    Cumulative share & 90\% of each intermediary\textquotesingle s traffic (union)          & 0.0066 & 0.0235 & 3.58 & 3.60 \\
    \midrule
    Minimum referrals & $\geq 5$ from each intermediary & 0.0111 & 0.0529 & 4.76 & 4.86 \\
    Minimum referrals & $\geq 10$ from each intermediary & 0.0140 & 0.0636 & 4.54 & 4.69 \\
    Minimum referrals & $\geq 50$ from each intermediary & 0.0260 & 0.0958 & 3.69 & 4.05 \\
    Minimum referrals & $\geq 100$ from each intermediary & 0.0353 & 0.1118 & 3.17 & 3.61 \\
    \midrule
    Classified domains & 4{,}266-domain classification universe   & 0.0134 & 0.0355 & 2.64 & 2.68 \\
    \midrule
    All domains (main) & All domains & 0.0055 & 0.0191 & 3.45 & 3.47 \\
    \bottomrule
  \end{tabular}
  \exhibitnotes{The full ten-month descriptive referral sample is restricted using each row\textquotesingle s rule. The ChatGPT and Google columns report raw Herfindahl indices computed from shares within each set. The ratio columns divide Google\textquotesingle s index by ChatGPT\textquotesingle s, using raw and normalized indices, respectively. Normalization uses each intermediary\textquotesingle s observed destination count within that set. The main paper reports normalized levels, so its levels differ from the raw columns here. No classifier-confidence restriction applies.}
\end{table}

\subsection{Concentration within household-weeks}\label{oa:within-user-conc}

The household comparison asks how evenly one household spreads its referrals within a week. For each household-week we compute the normalized index separately for ChatGPT and Google, and we regress it on a ChatGPT indicator with household and week fixed effects (\autoref{tab:oa-within-user-conc}). Each column requires a minimum number of referrals from each intermediary, and the index requires at least two distinct destinations.

The difference is close to zero when one or two referrals are required. It is positive at higher minimums and reaches about 0.06 at ten or more. At those minimums a household's ChatGPT referrals are more concentrated than its Google referrals, while ChatGPT's referrals are less concentrated in aggregate. The two patterns are consistent with different households sending their ChatGPT referrals to different destinations.

\begin{table}[!htbp]
  \centering
  \caption{The within-household concentration difference depends on referral activity.}
  \label{tab:oa-within-user-conc}
  \footnotesize
  \setlength{\tabcolsep}{4pt}
  \begin{tabular}{lrrrrrrr}
  \toprule
  Minimum referrals & $\geq1$ & $\geq2$ & $\geq3$ & $\geq5$ & $\geq10$ & $\geq15$ & $\geq20$ \\
  \midrule
  ChatGPT indicator & $-0.001$ & $0.000$ & $0.019^{***}$ & $0.035^{***}$ & $0.056^{***}$ & $0.065^{***}$ & $0.063^{***}$ \\
   & (0.002) & (0.002) & (0.002) & (0.003) & (0.006) & (0.012) & (0.018) \\
  ChatGPT mean HHI & 0.069 & 0.070 & 0.091 & 0.107 & 0.130 & 0.150 & 0.159 \\
  Google mean HHI & 0.072 & 0.071 & 0.074 & 0.075 & 0.077 & 0.088 & 0.099 \\
  \midrule
  Household fixed effects & Yes & Yes & Yes & Yes & Yes & Yes & Yes \\
  Week fixed effects & Yes & Yes & Yes & Yes & Yes & Yes & Yes \\
  Observations & 48{,}641 & 35{,}644 & 24{,}553 & 12{,}684 & 3{,}721 & 1{,}541 & 785 \\
  \bottomrule
  \end{tabular}
  \exhibitnotes{The dependent variable is household-week normalized HHI. Each specification requires the stated minimum referrals from both intermediaries and excludes household-week-intermediary observations with fewer than two destinations, for which the index is undefined. Household and week fixed effects are included throughout. Positive ChatGPT coefficients indicate greater concentration than Google. Means describe the estimation sample. Standard errors are clustered by household; $^{***}p<0.01$.}
\end{table}

\subsection{Destination prominence}\label{oa:robots}

\begin{figure}[!htbp]
  \centering
  \caption{ChatGPT allocates relatively larger referral shares to less-prominent destinations.}
  \label{fig:oa-authority-tier}

  \includegraphics[width=0.72\linewidth]{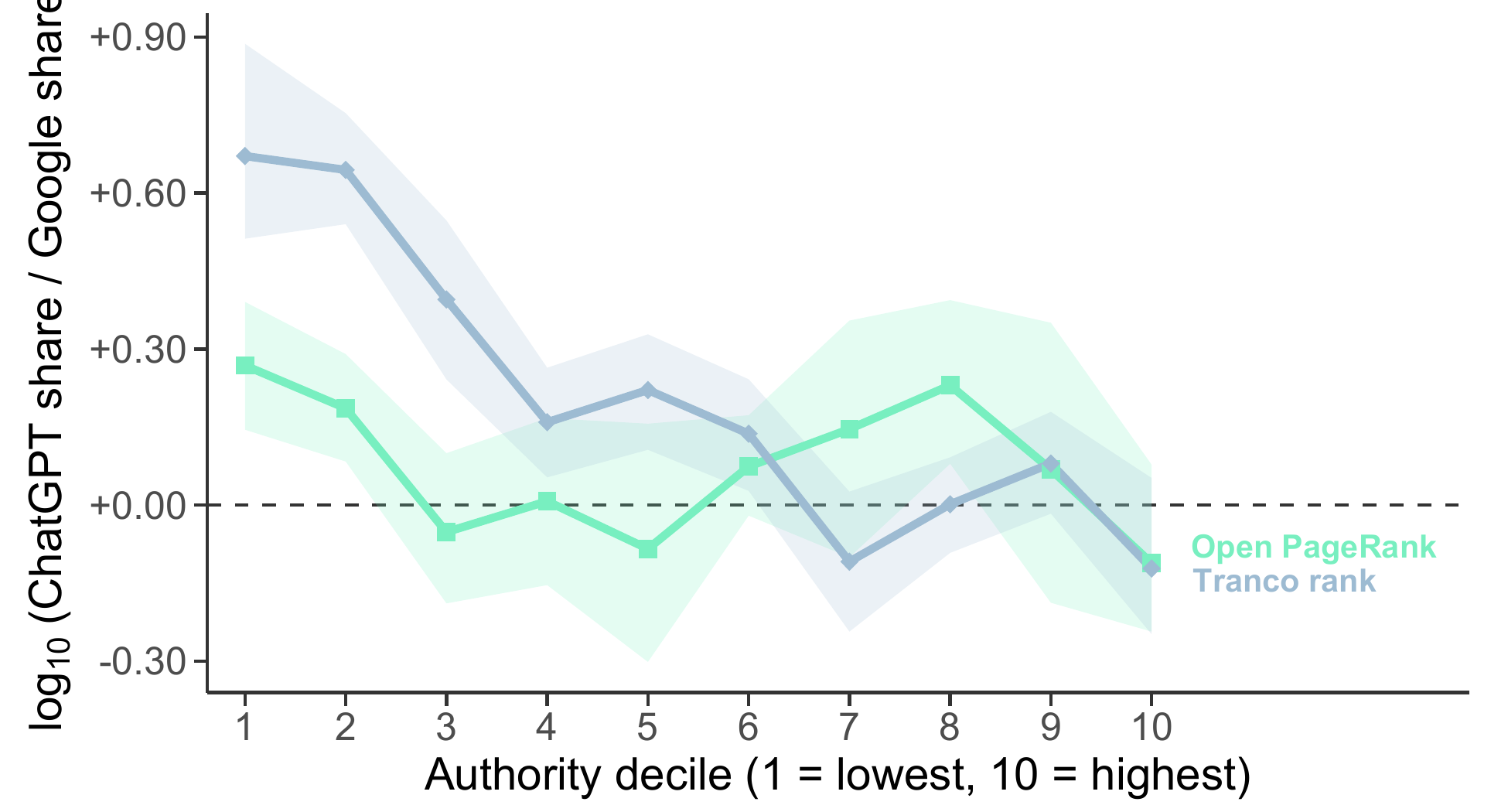}

  \exhibitnotes{Among the classified destinations with observed referral shares and available ranks, 4{,}201 enter the Open PageRank series and 3{,}959 enter the Tranco series. Each ranking separately groups these domains into deciles, from least prominent (one) to most prominent (ten); the axis labels these deciles as authority. The vertical axis is the base-ten logarithm of the ChatGPT share divided by the Google share within each decile. Positive values indicate a relatively larger ChatGPT share. Shaded bands are pointwise 95\% percentile intervals from 1{,}000 bootstrap resamples of domains within each ranking and decile.}
\end{figure}

ChatGPT sends a relatively larger share of its referrals to less prominent websites than Google does (\autoref{fig:oa-authority-tier}). The figure groups websites into deciles of two public rankings, Open PageRank and Tranco, which measure a website's prominence by its links and its popularity.

\section{Sustained adoption and conventional search}\label{oa:causal}

Main-paper \autoref{sec:displacement} asks how conventional search changes when households use ChatGPT, and it uses two comparisons. The first, documented in this section, follows households as they begin sustained ChatGPT use and compares them with households that adopt later. The second, in \autoref{oa:access}, compares households that gained Search access on February~5, 2025 with households that used Gemini. Both use the balanced panel of \panelN{} households and the measures and weights below.

\subsection{Measures and demographic weighting}\label{oa:acs-reweighting}

\paragraph{Conventional searches.} Conventional searches are a household's weekly number of Google, Bing, and Yahoo search-result page loads: foreground visits (\autoref{oa:cleaning}) whose URL path ends in \path{/search}. Each load counts once, so the measure counts page loads and not distinct questions. A week with no loads counts as zero.

\paragraph{ChatGPT use.} ChatGPT use is the weekly number of ChatGPT sessions (\autoref{oa:conv-cleaning}).

\paragraph{Weights.} Both comparisons weight households to match the U.S. population by age and income, using the 2023 American Community Survey (ACS), and they drop households without a positive weight. Weighted means normalize the weights within each group. The weights adjust the panel's composition by age and income; households may still differ from the population in other ways, and activity on mobile devices is outside the data.

\subsection{Sample and estimator}\label{oa:lbs-replication}

The design follows \citet{padilla2025}. A household adopts in the first of at least three consecutive complete weeks with a ChatGPT session. The three-week requirement marks the start of sustained use, and a household may have used ChatGPT occasionally before it. Weeks run Tuesday through Monday, which gives 43 complete weeks from October~1, 2024 through July~28, 2025.

Of 1{,}361 households with such a run, 282 begin in the first four weeks (October~1--28, 2024). We exclude them so that every adopter has at least four weeks of pre-adoption history. That leaves 1{,}079 adopters, of whom 1{,}008 have positive ACS weights. Our setting differs from \citet{padilla2025} in three ways. We study ChatGPT alone, where they study several LLMs. We define the panel by four visits per month, where they require four active days. And our data end in July~2025, so later adoption is unobserved.

Each adoption cohort is compared with households that have not yet adopted. For cohort $g$ and calendar week $t$, these are the adopters whose adoption week $G_i$ satisfies $G_i>\max(g,t)$, and the estimate is the unconditional group-time estimator of \citet{callaway2021}:
\begin{equation}
\widehat{ATT}(g,t)=\overline{Y_{it}-Y_{i,g-1}}_{G_i=g,\,w}-\overline{Y_{it}-Y_{i,g-1}}_{G_i>\max(g,t),\,w},
\label{eq:oa-adoption}
\end{equation}
where the subscript $w$ marks ACS-weighted means. An estimate requires at least two adopting and two comparison households. We aggregate the estimates by weeks since adoption, weighting cohorts by their ACS weight, and we average them over windows of weeks with each week weighted equally. Standard errors are clustered by household and account for the estimated cohort weights and the correlation across weeks. The estimates have a causal reading if searches in adopting and not-yet-adopting households would have followed parallel trends without adoption, and if households did not change their search in anticipation of adopting.

The joint test of the pre-adoption estimates does not reject at the 5\% level ($p=0.295$; \autoref{tab:oa-pretrend}, Panel A).

\subsection{Earlier and later adopters}\label{oa:adoption-gmm}

Main-paper \autoref{fig:adoption-timing} splits adopters at February~5, 2025, the day Search became available without an account, using the first session day of the adoption week. There are 366 earlier and 642 later adopters. The 45 households in the last adoption cohort have no later adopters to be compared with and serve only as comparisons, which leaves 597 later adopters with estimates. Later adopters are observed for fewer weeks: 54 of them are observed at week 19, the last week reported for that group. Pre-adoption tests for the two groups do not reject either ($p=0.280$ for earlier and $0.291$ for later adopters).

The quantity of interest is the later-minus-earlier contrast at each week $w$ since adoption, $\widehat{\Delta}_w=\widehat{ATT}_{\mathrm{later},w}-\widehat{ATT}_{\mathrm{earlier},w}$. It compares the two groups at the same number of weeks since adoption. The groups adopt in different calendar periods and can differ in other ways, so the contrast reflects those differences together with Search availability.

\paragraph{Four-baseline GMM.} Main-paper \autoref{fig:adoption-timing}(c) and the final column of \autoref{tab:adoption-windows} estimate the contrast with a generalized method of moments (GMM) estimator that uses four pre-adoption weeks as baselines, where \autoref{eq:oa-adoption} uses one. For adopter group $s$, cohort $g$, and week $t$, let $\widehat d_{s,g,t,b}$ be the estimate in \autoref{eq:oa-adoption} with baseline week $b\in\mathcal B_g=\{g-4,g-3,g-2,g-1\}$, computed for the adopters in group $s$ against the same comparison households. Every adopter has these four weeks because the sample requires four weeks of history. Stack the four estimates in $\widehat{\boldsymbol d}_{s,g,t}$ and let $\widehat V_{s,g,t}$ be their covariance across households. The combined estimate is
\begin{equation}
\widehat\tau^{\mathrm{GMM}}_{s,g,t}
=\widehat{\boldsymbol a}_{s,g,t}'\widehat{\boldsymbol d}_{s,g,t},
\qquad
\widehat{\boldsymbol a}_{s,g,t}
=\frac{\widehat V_{s,g,t}^{-1}\boldsymbol 1}
{\boldsymbol 1'\widehat V_{s,g,t}^{-1}\boldsymbol 1}.
\label{eq:oa-adoption-gmm}
\end{equation}
The weights sum to one and can be negative. Combining baselines improves precision when the four estimates measure the same effect \citep{chen2025efficient}. We aggregate over cohorts by ACS weight, as before, and subtract the earlier group's estimate from the later group's. Outcomes are weekly search counts without winsorizing. The assumptions are parallel trends from each of the four baselines and no anticipation of adoption.

Inference uses a household bootstrap with 999 draws. Each draw reweights households with independent $\operatorname{Exp}(1)$ weights and re-estimates every step, including the covariance weights and the cohort shares. We report pointwise 95\% basic bootstrap intervals and two-sided bootstrap tests for the window averages. Estimates for pre-adoption weeks omit a baseline that coincides with the week being estimated, so week $-1$ is estimated from weeks $-4$ through $-2$ and is not set to zero.

Over weeks 4--19 the contrast averages $-13.01$ searches per household-week ($SE=3.91$; 95\% interval $[-20.68,-5.01]$). Three diagnostics qualify this estimate. The joint test of the estimates for weeks $-12$ through $-1$ rejects at the 5\% level ($p=0.023$). Some cohorts contain only five adopting households, and the baseline weights range from $-16.91$ to $10.30$. Dropping any one household leaves the window average between $-13.93$ and $-11.00$, though estimates for single weeks move more. We chose GMM after comparing estimators, so this inference is exploratory.

\begin{figure}[!htbp]
\centering
\caption{The single-baseline contrast also shows conventional searches declining sooner among later adopters.}
\label{fig:oa-adoption-original-contrast}
\begin{minipage}[c]{0.04\linewidth}
\centering\rotatebox{90}{\small Weekly conventional searches}
\end{minipage}%
\begin{minipage}[c]{3.053in}
\centering
\includegraphics[width=\linewidth]{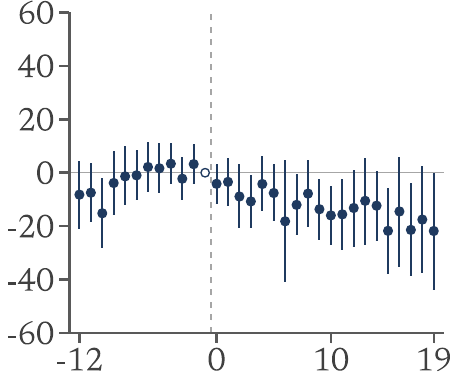}
\par\smallskip\small Weeks from adoption
\end{minipage}
\exhibitnotes{Points show the later-minus-earlier difference in the estimated change in weekly Google, Bing, and Yahoo searches per household. Estimates are ACS-weighted, use not-yet-adopting households as comparisons, and take week $-1$ as the reference. The sample has 366 earlier and 642 later adopters, 45 of whom serve only as comparisons; the number of later adopters with estimates falls from 597 at week 0 to 54 at week 19. Error bars are pointwise 95\% bootstrap intervals from 9{,}999 household-level draws. Main-paper \autoref{fig:adoption-timing}(c) shows the four-baseline GMM contrast.}
\end{figure}

\paragraph{Single-baseline comparison.} The contrast is similar when each group is estimated with the single baseline of \autoref{eq:oa-adoption}, as in panels (a) and (b) of main-paper \autoref{fig:adoption-timing}. Subtracting panel (a) from panel (b) gives an average of $-14.26$ over weeks 4--19 ($SE=5.27$; \autoref{fig:oa-adoption-original-contrast}). Its intervals come from a multiplier bootstrap of the household influence functions of \citet{callaway2021} with 9{,}999 draws. Each draw applies one multiplier per household to both groups and all weeks, which preserves the dependence created by shared comparison households.

\paragraph{Fixed sample of later adopters.} Part of the contrast reflects which later adopters are observed at each week. Over weeks 0--19 the contrast is $-12.77$ ($SE=4.60$) when each week uses the adopters observed in it. Holding the later group fixed at the 54 households observed through week 19 gives $-7.51$ ($SE=4.39$; \autoref{tab:oa-adoption-alternatives}). Much of the faster decline among later adopters therefore comes from households that began sustained use from late February onward and are observed into the summer. Unweighted estimates, which add the 71 adopters without ACS weights, are close to the weighted ones.

\begin{table}[!htbp]
\centering
\caption{Holding the sample of later adopters fixed narrows the difference between groups.}
\label{tab:oa-adoption-alternatives}
\footnotesize
\setlength{\tabcolsep}{4pt}
\begin{tabular}{lrrr}
\toprule
Weeks 0--19 & Earlier & Later & Later minus earlier \\
\midrule
\multicolumn{4}{l}{\textit{Panel A. ACS-weighted}} \\
\addlinespace
All available adopters & 2.18 (2.64) & -10.59 (3.90) & -12.77 (4.60) \\
Adopters observed through week 19 & 2.18 (2.64) & -5.33 (3.66) & -7.51 (4.39) \\
\midrule
\multicolumn{4}{l}{\textit{Panel B. Unweighted}} \\
\addlinespace
All available adopters & 3.67 (2.29) & -8.58 (4.01) & -12.25 (4.63) \\
Adopters observed through week 19 & 3.67 (2.29) & -2.42 (4.50) & -6.08 (5.03) \\
\bottomrule
\end{tabular}
\par\smallskip\begin{minipage}{\linewidth}\footnotesize \textit{Notes:} Entries are changes in weekly conventional searches averaged over weeks 0--19 after sustained adoption, with household-clustered standard errors in parentheses. Earlier adopters adopt before February 5, 2025 and later adopters on or after that date, using the first session day of the adoption week. In each panel, the first row uses the adopters observed in each week, and the second row uses the adoption cohorts observed through week 19, so the same cohorts contribute to every week. ACS weights adjust for age and income. Standard errors account for shared comparison households and estimated cohort shares.\end{minipage}
\end{table}

\section{The February access expansion and conventional search}\label{oa:access}

On February~5, 2025, ChatGPT Search became available to users without an account. The second comparison in main-paper \autoref{sec:displacement} uses this date. It compares households that had been using ChatGPT without an account, and so gained Search on that day, with households that had been using Gemini. This section defines the two groups, the estimator, and the checks. It then reports GMM estimates and the referral outcomes by website category. Measures and weights are those of \autoref{oa:acs-reweighting}.

\subsection{Treated and comparison households}\label{oa:cohort-schema}

Group membership is fixed using activity before February~5, so a household's later behavior cannot change its group.

\paragraph{Treated households.} A treated household used ChatGPT only without an account before the expansion. It has at least three ChatGPT requests before February~5, at least one day of anonymous use, no day of logged-in or mixed use, and no request for a paid feature. The data contain no login or subscription records, so we infer how a household used ChatGPT from its request paths: the application sends requests under \texttt{backend-anon} for anonymous users and under \texttt{backend-api} for logged-in users (\autoref{tab:oa-endpoint-defs}). A day counts as anonymous or logged-in use according to its successful requests on the conversation paths, and requests to the account and conversation-list paths also mark logged-in use. There are 946 such households.

\paragraph{Comparison households.} Comparison households used Gemini before February~5 and had no ChatGPT, OpenAI, Claude, or Anthropic activity before the expansion, including no foreground ChatGPT visit between October~1 and February~4. They stay in the comparison group if they use ChatGPT later. We keep households in demographic categories that appear in both groups, across age, income, household size, children, gender, race, and location. This drops one treated household with unknown household size and leaves 945 treated and 429 comparison households. With ACS weights, the effective sample sizes, $(\sum_i w_i)^2/\sum_i w_i^2$, are 774 and 355.

\paragraph{Prior search activity.} We split households by how much they searched before the expansion. For each household we average weekly conventional searches over the 18 pre-expansion weeks (October~2, 2024--February~4, 2025), after capping each week at 66 searches, the 95th percentile of weekly searches in the 30{,}757-household sample from which the two groups are drawn. Households at or below the treated group's ACS-weighted median of 18.5 form the lower group (492 treated, 202 comparison), and those above it form the higher group (453 treated, 227 comparison).

\subsection{Estimation and inference}\label{oa:master-table}

Event week 0 begins on February~5, 2025. Weeks run Wednesday through Tuesday, and we use weeks $-18$ through 24, which end on July~29. Weekly searches are winsorized at 166, the unweighted 99th percentile of household-week searches, zeros included, in the 30{,}757-household sample over the 18 pre-expansion weeks. ChatGPT sessions and category referrals are not capped.

The main estimates compare the treated-minus-comparison gap in week $k$ with the gap in week $-1$ (January~29--February~4). For outcome $Y$ and treated indicator $T_i$,
\begin{equation}
\widehat\delta_k=\bigl(\bar Y_{T=1,k}-\bar Y_{T=0,k}\bigr)-\bigl(\bar Y_{T=1,-1}-\bar Y_{T=0,-1}\bigr),
\label{eq:oa-event}
\end{equation}
where bars denote ACS-weighted means. Because the panel is balanced and the weights are fixed, these are the coefficients of an event-study regression with household and calendar-week fixed effects (TWFE). Main-paper \autoref{tab:february-access} and \autoref{tab:prior-search} report their averages over weeks 0--24 for all households and for the two prior-search groups. The comparison identifies the effect of the expansion if, without it, the gap between the two groups would have stayed at its week $-1$ level.

Standard errors come from a household bootstrap with 999 draws. Each draw reweights households with independent $\operatorname{Exp}(1)$ weights and re-estimates the means. Group membership, the search cap, and the ACS weights are held at their sample values. Intervals are pointwise 95\% basic bootstrap intervals, without adjustment for the number of weeks. We settled on the estimator and the prior-search split after comparing alternatives, and the bootstrap does not account for that choice. Inference for this design, including the category estimates in \autoref{oa:category-outcomes}, is therefore exploratory.

\subsection{Pre-expansion trends and weekly responses}\label{oa:pretrend}

The treated and comparison groups followed similar paths before the expansion. Joint tests of the pre-expansion estimates do not reject at the 5\% level in the full sample, for searches ($p=0.248$) or for sessions ($p=0.433$), or in either prior-search group (\autoref{tab:oa-pretrend}, Panel B). Failing to reject does not establish that trends were parallel, and a test restricted to the last weeks before the expansion, $-9$ through $-2$, has $p=0.068$ in the full sample.

\begin{table}[!htbp]
\centering
\caption{Pre-period tests for the single-baseline specifications do not reject at the 5\% level.}
\label{tab:oa-pretrend}
\footnotesize
\setlength{\tabcolsep}{5pt}
\begin{tabular}{llrrr}
\toprule
Group & Weighting & Search $p$ & Late search $p$ & Session $p$ \\
\midrule
\multicolumn{5}{l}{\textit{Panel A. Sustained adoption}} \\
\addlinespace
All adopters & ACS & $0.295$ & --- & --- \\
Earlier adopters & ACS & $0.280$ & --- & --- \\
Later adopters & ACS & $0.291$ & --- & --- \\
\addlinespace
All adopters & Unweighted & $0.238$ & --- & --- \\
Earlier adopters & Unweighted & $0.458$ & --- & --- \\
Later adopters & Unweighted & $0.109$ & --- & --- \\
\midrule
\multicolumn{5}{l}{\textit{Panel B. February access: Gemini comparison}} \\
\addlinespace
All households & ACS & $0.248$ & $0.068$ & $0.433$ \\
Lower prior search & ACS & $0.293$ & $0.135$ & $0.408$ \\
Higher prior search & ACS & $0.405$ & $0.212$ & $0.417$ \\
\bottomrule
\end{tabular}
\exhibitnotes{Entries are $p$-values from joint Wald tests that the pre-period estimates are zero, using household-clustered covariance. Panel A tests conventional-search estimates for weeks $-12$ through $-2$, relative to the week before sustained adoption; earlier adopters adopt before February 5, 2025. Panel B uses the Gemini comparison with week $-1$ as reference: the search and session tests cover weeks $-18$ through $-2$, and the late-search test covers weeks $-9$ through $-2$. February conventional searches are winsorized at 166 per household-week; sessions are not capped. ACS denotes age-by-income weights. Dashes indicate tests not reported.}
\end{table}

\begin{figure}[!htbp]
\centering
\caption{Conventional-search declines concentrate among households with lower prior activity.}
\label{oa:prior-search-event}
\begin{minipage}[t]{0.49\linewidth}
\centering
\includegraphics[width=\linewidth]{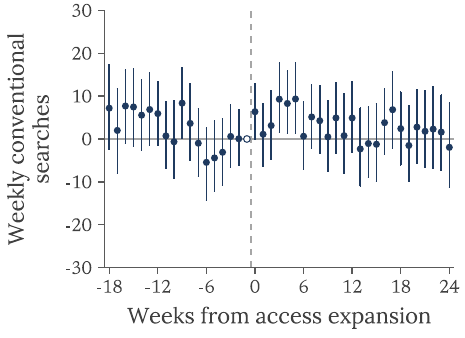}
\smallskip\small (a) Higher prior search activity.
\end{minipage}\hfill
\begin{minipage}[t]{0.49\linewidth}
\centering
\includegraphics[width=\linewidth]{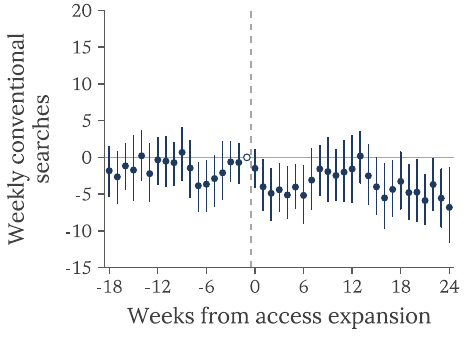}
\smallskip\small (b) Lower prior search activity.
\end{minipage}
\exhibitnotes{Points show ACS-weighted TWFE estimates of effects on weekly Google, Bing, and Yahoo searches, with household and calendar-week fixed effects. Panel (a) compares 453 households that used ChatGPT without an account before the expansion with 227 households that used Gemini; panel (b) compares 492 with 202. Groups use mean conventional searches over all 18 pre-expansion weeks (\autoref{oa:cohort-schema}). Open circles mark week $-1$, normalized to zero. Bars show pointwise 95\% basic intervals from 999 household bootstrap draws. Dashed lines mark the February~5 expansion. Panels use different vertical scales.}
\end{figure}

\autoref{oa:prior-search-event} shows the weekly estimates behind main-paper \autoref{tab:prior-search}. Conventional searches decline in the lower-activity group and show no detectable change in the higher-activity group.

\subsection{Comparison with GMM}\label{oa:model-specification}\label{oa:access-gmm}

The main estimates use one reference week. As a check, we combine all 18 pre-expansion weeks as baselines with GMM \citep{chen2025efficient}, the approach used for the adoption contrast in \autoref{oa:adoption-gmm}. For each outcome and sample, let $\widehat d_{k,b}$ be the ACS-weighted change in the treated-minus-comparison gap from baseline week $b\in\mathcal B=\{-18,\ldots,-1\}$ to week $k$. Stack these changes in $\widehat{\boldsymbol d}_k$ and let $\widehat V_k$ be their covariance across households. The estimate is
\begin{equation}
\widehat\delta^{\mathrm{GMM}}_k=
\frac{\boldsymbol 1'\widetilde V_k^{-1}\widehat{\boldsymbol d}_k}
{\boldsymbol 1'\widetilde V_k^{-1}\boldsymbol 1},
\qquad \widetilde V_k=\widehat V_k+\eta_k I,
\label{eq:oa-access-gmm}
\end{equation}
where $\eta_k$ is a small constant that keeps $\widetilde V_k$ invertible when session counts are sparse. The weights sum to one and can be negative, and they are estimated separately for each outcome, sample, and week. Estimates for pre-expansion weeks leave out the baseline that coincides with the week being estimated, so week $-1$ is estimated and not set to zero. Post-expansion effects average weeks 0--24. The assumptions are parallel trends from every baseline week and no anticipation of the expansion.

GMM leads to the same conclusions as the main estimates (\autoref{tab:oa-access-gmm}). ChatGPT use rises by 0.127 sessions per household-week ($SE=0.020$), and conventional searches show no detectable overall change (0.21, $SE=1.02$). Searches fall by 2.81 per week in the lower-activity group ($SE=0.78$) and rise by an imprecise 2.47 in the higher-activity group ($SE=1.75$). The lower-minus-higher contrast is $-5.29$ ($SE=1.91$).

GMM is also somewhat more precise for searches. A natural benchmark weights all pre-expansion weeks equally, as in the ACS-weighted regression
\begin{equation}
Y_{it}=\alpha_i+\lambda_t+\beta\,T_i\,\mathbf{1}\{t\geq0\}+\varepsilon_{it},
\label{eq:oa-static}
\end{equation}
whose coefficient $\beta$ is the average post-expansion gap minus the average pre-expansion gap. Relative to that benchmark and with the same bootstrap draws, GMM lowers the standard error of the search estimate from 1.16 to 1.02 in the full sample, from 0.93 to 0.78 in the lower-activity group, and from 2.00 to 1.75 in the higher-activity group, a gain of about 12--16\%. Session estimates are about equally precise, and slightly less precise in the lower-activity group.

\begin{table}[!htbp]
\centering
\caption{GMM estimates of the access expansion are close to the main estimates.}
\label{tab:oa-access-gmm}
\footnotesize
\setlength{\tabcolsep}{5pt}
\begin{tabular}{lrrr}
\toprule
 & All households & Lower prior search & Higher prior search \\
\midrule
ChatGPT sessions & $0.127^{***}$ & $0.105^{***}$ & $0.151^{***}$ \\
 & (0.020) & (0.024) & (0.032) \\
\addlinespace
Conventional searches & $0.21$ & $-2.81^{***}$ & $2.47$ \\
 & (1.02) & (0.78) & (1.75) \\
\addlinespace
\midrule
Treated / comparison & 945 / 429 & 492 / 202 & 453 / 227 \\
TWFE conventional-search lead-test $p$ & 0.248 & 0.293 & 0.405 \\
Lower $-$ higher conventional searches &  & $-5.29^{***}$ &  \\
 &  & (1.91) &  \\
\bottomrule
\end{tabular}
\par\smallskip\begin{minipage}{\linewidth}\footnotesize \textit{Notes:} GMM estimates combine all 18 pre-expansion weeks as baselines (\autoref{oa:access-gmm}) and use the same households, ACS weights, prior-search groups, and outcomes as main-paper \autoref{tab:february-access} and \autoref{tab:prior-search}. Estimates average weeks 0--24 after February~5, 2025. Conventional searches (Google, Bing, and Yahoo result-page loads) are winsorized at the 99th percentile. The lead-test row reports the TWFE test of conventional-search estimates in weeks $-18$ to $-2$, relative to week $-1$. Standard errors from 999 household bootstrap draws are in parentheses. $^{*}p<0.10$, $^{**}p<0.05$, $^{***}p<0.01$.\end{minipage}
\end{table}

\subsection{Search referrals by category}\label{oa:category-outcomes}

The last set of outcomes asks which kinds of websites lose or gain visits referred by conventional search after the expansion.

\paragraph{Outcome.}\label{oa:referral-measures} The outcome is a household's weekly number of visits referred by Google, Bing, or Yahoo to websites in a content category. A visit is a clean foreground visit to a destination page (\autoref{oa:referral-def}), and each counts once. Referrals are counted from the referrer alone and are not assigned to individual queries. Categories use the 3{,}103 domains classified with confidence of at least 0.90 in the eleven content categories that remain after excluding other and unknown (\autoref{oa:classification}). Counts are not capped, and a household-week with no such visits counts as zero. The households, weights, and weeks are the same as for the main outcomes (\autoref{oa:cohort-schema} and \autoref{oa:master-table}): 945 treated and 429 comparison households.

\paragraph{Baseline composition.} Panel~(a) of main-paper \autoref{fig:prior-search-composition} describes where conventional-search referrals led before the expansion, among treated households. For prior-search group $g$ and category $c$, the share is
\begin{equation}
s_{gc}=\frac{\sum_{i\in g}w_i R^{\mathrm{pre}}_{ic}}{\sum_{i\in g}w_i\sum_{k=1}^{11}R^{\mathrm{pre}}_{ik}},
\end{equation}
where $R^{\mathrm{pre}}_{ic}$ counts household $i$'s referred visits to category $c$ over the 18 pre-expansion weeks and $w_i$ is its ACS weight. The lower-activity group sends a larger share of its referrals to tools/SaaS websites, by 4.5 percentage points (95\% interval [2.6, 6.4]; adjusted $p<0.001$), and a smaller share to social media, by 9.8 points ([$-14.7,-4.9$]; adjusted $p<0.001$). Adult websites receive 3.2\% of the lower-activity group's referrals and 6.7\% of the higher-activity group's, a difference of $-3.5$ points ([$-5.9,-1.0$]; adjusted $p=0.036$). Reference/knowledge shares are similar, 18.8\% and 16.9\%. Standard errors treat each share as a ratio and are clustered by household, and $p$-values are Holm-adjusted across the eleven categories.

\paragraph{Category effects.} For each category we estimate the TWFE specification in \autoref{eq:oa-event}, average weeks 0--24, and express the result as a percentage of the treated group's mean for that category over the 18 pre-expansion weeks. Panel~(b) of main-paper \autoref{fig:content-heterogeneity} reports these estimates for all eleven categories, including adult websites. Its pointwise 95\% intervals use household-clustered standard errors and normal critical values, and they treat the pre-expansion mean as fixed.

\paragraph{TWFE and GMM.} The category estimates are sensitive to the estimator, so they give weaker support for conclusions about particular uses of search than the overall and prior-search results do (\autoref{tab:oa-categories}). In the full sample, the tools/SaaS estimate is $-46.0\%$ under TWFE and $-10.0\%$ under GMM, the e-commerce brand estimate is $-26.5\%$ and $-9.7\%$, and the social media estimate is $-11.0\%$ and $2.1\%$. All ten full-sample GMM intervals include zero.

The table covers ten categories and omits adult websites. Its full-sample TWFE estimates equal those in main-paper \autoref{fig:content-heterogeneity}(b). The table's significance marks come from bootstrap tests and the figure's from analytic standard errors, so the marks can differ.

Pre-expansion trends also differ across categories under TWFE. The joint test of pre-expansion estimates does not reject for tools/SaaS ($p=0.377$), e-commerce brands ($p=0.326$), or social media ($p=0.930$). It rejects for reference/knowledge ($p=0.009$), whose $-39.5\%$ TWFE estimate has a 95\% interval excluding zero, and for news/journalism ($p=0.016$) and government/institutional ($p=0.029$).

For the three categories that the main paper discusses, the GMM estimates within prior-search groups are a decline of 12.8\% in referrals to tools/SaaS and of 24.6\% in referrals to e-commerce brands in the lower-activity group, and an increase of 3.6\% in referrals to social media in the higher-activity group. Of these three, only the e-commerce brand interval excludes zero, and its unadjusted $p=0.030$ becomes $0.270$ after a Holm correction across the ten categories.

\begin{table}[!htbp]
\centering
\caption{Category referral effects are sensitive to the estimator.}
\label{tab:oa-categories}
\footnotesize
\setlength{\tabcolsep}{4pt}
\begin{tabular}{lrrrrrr}
\toprule
 & \multicolumn{2}{c}{All households} & \multicolumn{2}{c}{Lower prior search} & \multicolumn{2}{c}{Higher prior search} \\
\cmidrule(lr){2-3} \cmidrule(lr){4-5} \cmidrule(lr){6-7}
Destination category & TWFE & GMM & TWFE & GMM & TWFE & GMM \\
\midrule
Academic research & $-20.1$ & $-8.1$ & $-21.5$ & $6.2$ & $-20.2$ & $-8.5$ \\
 & (21.8) & (6.7) & (27.9) & (14.7) & (25.3) & (6.9) \\
\addlinespace[2pt]
Developer/technical & $30.2$ & $-9.4$ & $-98.1^{*}$ & $-25.5$ & $62.8$ & $-0.2$ \\
 & (75.9) & (12.7) & (61.7) & (19.7) & (91.6) & (14.2) \\
\addlinespace[2pt]
E-commerce brand & $-26.5^{*}$ & $-9.7$ & $-44.2^{*}$ & $-24.6^{**}$ & $-21.0$ & $-5.4$ \\
 & (15.8) & (6.4) & (24.6) & (10.8) & (18.8) & (7.4) \\
\addlinespace[2pt]
E-commerce marketplace & $3.0$ & $6.0$ & $-46.7$ & $-19.4$ & $12.4$ & $15.8^{**}$ \\
 & (16.8) & (8.5) & (31.2) & (27.6) & (18.3) & (7.4) \\
\addlinespace[2pt]
Games and guides & $-15.9$ & $1.8$ & $-29.8$ & $5.9$ & $-13.2$ & $0.9$ \\
 & (19.5) & (9.7) & (38.5) & (19.3) & (21.0) & (10.5) \\
\addlinespace[2pt]
Government/institutional & $-4.0$ & $8.6$ & $16.2$ & $9.8$ & $-11.9$ & $2.9$ \\
 & (20.6) & (7.9) & (34.6) & (13.8) & (24.5) & (9.1) \\
\addlinespace[2pt]
News/journalism & $-32.5$ & $0.6$ & $-30.2$ & $21.4^{**}$ & $-33.8$ & $-6.0$ \\
 & (21.7) & (7.7) & (28.9) & (10.6) & (23.9) & (8.1) \\
\addlinespace[2pt]
Reference/knowledge & $-39.5^{**}$ & $1.5$ & $-71.4^{*}$ & $-29.9^{***}$ & $-33.1^{*}$ & $2.2$ \\
 & (14.5) & (5.7) & (32.6) & (8.8) & (15.6) & (6.3) \\
\addlinespace[2pt]
Social media & $-11.0$ & $2.1$ & $-40.0$ & $-13.4$ & $-8.4$ & $3.6$ \\
 & (14.0) & (4.9) & (29.8) & (12.2) & (15.1) & (5.2) \\
\addlinespace[2pt]
Tools/SaaS & $-46.0^{*}$ & $-10.0$ & $-37.7$ & $-12.8$ & $-49.8$ & $-9.6$ \\
 & (23.3) & (6.5) & (30.2) & (10.6) & (28.8) & (7.4) \\
\addlinespace[2pt]
\midrule
Treated households & 945 & 945 & 492 & 492 & 453 & 453 \\
Comparison households & 429 & 429 & 202 & 202 & 227 & 227 \\
\bottomrule
\end{tabular}
\par\smallskip\begin{minipage}{\linewidth}\footnotesize \textit{Notes:} ACS-weighted estimates compare households that used ChatGPT without an account before the expansion with households that used Gemini. Outcomes are uncapped weekly visits referred by Google, Bing, and Yahoo, averaged over weeks 0--24 after February~5, 2025. TWFE uses household and calendar-week fixed effects with week $-1$ as reference; GMM combines all 18 pre-expansion weeks as baselines. Estimates and standard errors are percentages of the treated group's mean for the category over the 18 pre-expansion weeks. Standard errors from 999 household bootstrap draws are in parentheses, and stars use unadjusted bootstrap tests. $^{*}p<0.10$, $^{**}p<0.05$, $^{***}p<0.01$.\end{minipage}
\end{table}

\end{document}